\documentclass[fleqn,usenatbib]{mnras}

\usepackage{newtxtext,newtxmath}

\usepackage[T1]{fontenc}
\usepackage{booktabs,siunitx}

\DeclareRobustCommand{\VAN}[3]{#2}
\let\VANthebibliography\thebibliography
\def\thebibliography{\DeclareRobustCommand{\VAN}[3]{##3}\VANthebibliography}

\usepackage{graphicx}	
\usepackage{amsmath}	
\usepackage{xcolor}
\newcommand{\dd}{{\rm d}}
\newcommand{\revise}[1]{#1}

\usepackage{hyperref}
\hypersetup{colorlinks, citecolor=blue}

\title[Precession of magnetars]{Precession of magnetars: dynamical evolutions
and modulations on polarized electromagnetic waves}

\author[Yong Gao et al.]{
Yong Gao,$^{1,2}$
Lijing Shao,$^{2,3,4}$\thanks{E-mail: lshao@pku.edu.cn (LS)}
Gregory Desvignes,$^{3,5}$
David Ian Jones,$^{6}$
Michael Kramer,$^{3,7}$
and Garvin Yim$^{2,6}$
\\
$^{1}$Department of Astronomy, School of Physics, Peking University, Beijing 100871, China\\
$^{2}$Kavli Institute for Astronomy and Astrophysics, Peking University, Beijing 100871, China\\
$^{3}$Max-Planck-Institut f\"ur Radioastronomie, Auf dem H\"ugel 69, D-53121 Bonn, Germany \\
$^{4}$National Astronomical Observatories, Chinese Academy of Sciences, Beijing 100012, China\\
$^{5}$Laboratoire d'\'Etudes Spatiales et d'Instrumentation en Astrophysique, Observatoire de Paris, Universit\'e Paris-Sciences-et-Lettres, \\ 
	  Centre National de la Recherche Scientifique, Sorbonne Universit\'e, Universit\'e de Paris, 5 place Jules Janssen, 92195 Meudon, France \\
$^{6}$Mathematical Sciences and STAG Research Centre, University of Southampton, Southampton SO17 1BJ, UK \\
$^{7}$Jodrell Bank Centre for Astrophysics, School of Physics and Astronomy, The University of Manchester, Manchester M13 9PL, UK
}

\date{Accepted XXX. Received YYY; in original form ZZZ}

\pubyear{2022}

\begin{document}
\label{firstpage}
\pagerange{\pageref{firstpage}--\pageref{lastpage}}
\maketitle

\begin{abstract}
Magnetars are conjectured to be highly magnetized neutron stars (NSs). Strong
internal magnetic field and elasticity in the crust may deform the stars and
lead to free precession. We study the precession dynamics of triaxially-deformed
NSs incorporating the near-field and the far-field electromagnetic torques.  We
obtain timing residuals for different NS geometries and torques. We also
investigate the polarized X-ray and radio signals from precessing magnetars. The
modulations on the Stokes parameters are obtained for  thermal X-rays emitted
from the surface of magnetars. For radio signals, we apply the simple rotating
vector model (RVM) to give the modulations on the position angle (PA) of the
polarization. Our results are comprehensive, ready to be used to search for
magnetar precession with timing data and polarizations of X-ray and radio
emissions. Future observations of precessing magnetars will give us
valuable information on the geometry and the strength of the strong magnetic
fields, the emission geometry, as well as the equation of state (EoS) of NSs.
\end{abstract}

\begin{keywords}
stars: magnetars -- methods: analytical -- X-rays: general -- polarization
\end{keywords}




\section{Introduction}

The free precession of neutron stars (NSs) has been studied since the discovery of
radio pulsars. The wobbling motion caused by the free precession is closely
related to the structure of NSs, such as the elasticity in the crust~
\citep{Ushomirsky:2000ax,Cutler:2002np}, the strength and the
geometry of the internal and external magnetic fields
\citep{Haskell:2007bh,Mastrano:2013jaa}, the superfluid and superconducting
states of the fluid interior
~\citep{Pines1974,Shaham1977,Sedrakian:1998vi,Link:2007zz,Glampedakis:2007hx},
as well as the evolution of the magnetic inclination
angle~\citep{Mestel1972,Goldreich:1970ads,Melatos:2000qt,Lander:2018und,lander2020}.

Although many theoretical works were devoted to this field, free precession has
not been firmly observed yet. The most probable evidence of free precession
comes from the radio pulsar PSR~B1828$-$11, which shows highly periodical
variations in the pulse phase over a period of $\sim 500\,\rm d$, accompanied by
correlated changes in the beam width of pulses~\citep{Stairs:2000zz}. The data
can be well fitted by the free precession model with the precession-modulated
spin-down torque~\citep{Jones:2000ud,Link:2001zr,Akgun:2005nd}. 
Meanwhile, radio observations started to show that several pulsars, 
with harmonic features in their timing residuals, including PSR~B1828$-$11, undergo sudden
sharp changes in pulse profiles which, at least for PSR~B1931+24, correlates with
sharp changes in spin-down torque~\citep{Kramer:2006ha,Lyne:2010ad,Shaw:2022mxv}.
\citet{Lyne:2010ad} argued that the radio emissions switch back and forth
between two different magnetospheric states, which leads to the harmonic timing
residuals and changes in the pulse shape. The sharpness of the transitions is
thought as strong evidence that free precession is not a viable mechanism to
explain the harmonic timing residuals in PSR~B1828$-$11. Further analysis by
\citet{Stairs:2019orm} also disfavored the free precession scenario. Mode
switching is very appealing and reasonable to explain the harmonic timing data
and shape variations, but the physics behind the sharp changes and the regular
clock of the transitions needs more development. \revise{\citet{Jones:2011jq} provided 
an argument as to why abrupt magnetospheric changes can occur in precessing stars.
In this regard, the free precession of PSR~B1828$-$11 cannot be ruled out firmly 
yet~\citep{,Ashton:2015wla,Ashton:2016vjx}.}

Different from normal pulsars, magnetars are a class of young NSs with strong
magnetic fields, typically above $10^{14}\,\rm G$, and long spin periods,
typically $2$--$10\,\rm s$~\citep{Kaspi:2017fwg}. 
The strong internal magnetic fields may distort the star~\citep{Melatos1997,Melatos:2000qt}.  
Due to the young age and energetic processes, the star may also develop large 
elastic deformation in the crust. Glitches or crust fracture may excite wobble angles and 
set the deformed magnetars into free precession. 
Strong magnetic fields also indicate large external electromagnetic torques, 
which include the far-field torque dissipating the kinetic energy and the near-field torque 
originating from the moment of inertia of the electromagnetic field 
itself~\citep{Goldreich:1970ads,Good1985,Melatos1997}. 
\revise{The forced precession can be significant for magnetars due to their strong electromagnetic torques, but this effect would
not affect the free precession if the object has even stronger internal magnetic fields reaching $10^{16}\,\rm G$~\citep{Makishima:2014dua}. We will 
investigate both free and forced precession in this work.}

Early observations of timing irregularities of magnetars motivated
\citet{Melatos1997} to suggest that precession is common in AXP populations.
However, further observations from X-ray timing in the band $\sim 1$--$10\,\rm
keV$ have ruled out precession at an amplitude level above the root-mean-square amplitude of 
timing noise~\citep{Kaspi:1999bj,Kaspi:2000fc}. Small-amplitude precession that is
buried in the timing noise of magnetars is still possible.
Possible evidence of magnetar precession were found by the combined timing
analysis of the hard and soft X-rays from three magnetars, 
4U~0142+61~\citep{Makishima:2014dua},
1E~1547.0$-$5408~\citep{Makishima:2021vvv}, and SGR~1900+14
\citep{Makishima:2021hho}. The phase modulations of hard X-ray are observed in
those magnetars, which may indicate large magnetic deformations on the order of 
$\sim 10^{-4}$ \citep{Makishima:2014dua,Makishima:2021vvv,Makishima:2021hho}.
Recently, the precession of magnetars was also used to interpret the possible
periodicity found in fast radio bursts
\citep[FRBs;][]{Levin:2020rhj,Zanazzi:2020vyp,Wasserman:2021dlh}.

All above evidences of magnetar precession come from the timing residuals, where 
the rotation phase is modulated by precession. During the precession, the body itself 
also precesses around the deformation axis, which leads to the swing of the emission region. 
The polarization directly maps the emission geometry. Thus it is also 
promising to reveal precession from the variations of polarizations. 

\revise{The soft component of the X-ray emission ($\sim
1$--$10\,\rm keV$) of magnetars is usually interpreted as
thermal emission from the magnetar surface, which is reprocessed by the strongly
magnetized atmosphere~\citep{Thompson:2001ig,Turolla:2015mwa}. 
Numerous studies have been dedicated to investigate the opacities and radiative transfer in strongly
magnetized atmospheres, showing that the surface emission could be highly 
polarized~\citep{Meszaros1992,Pavlov:1999cw,Ho:2002jn,Lai:2003nd,Taverna:2015vpa}.}
Recently, the IXPE observation of 4U~0142+61 gave the first ever measurement of
polarized emission from a magnetar in the soft X-ray
band~\citep{2022arXiv220508898T}.  The observations provided us completely new
information about the NS surface and magnetosphere and showed some evidence of 
vacuum birefringence in a strong magnetic field.

The radio emissions from magnetars are only observed in transient magnetars
after energetic bursts and the signal itself is also
transient~\citep{Camilo:2006zf,Camilo:2007jn,Levin:2010vm,Eatough2013,Lower:2020was}.
The emission is highly polarized.  By applying the rotating vector model
\citep[RVM;][]{rvm1969}, useful information was obtained on the magnetic field
geometry of magnetars~\citep{Kramer:2007fh,Levin:2012bi,Lower:2020dsz}.

In this paper, we aim to systematically study the dynamics of 
precessing magnetars and model the observational consequences in timing and 
polarization of electromagnetic waves. The
structure of the paper is organized as follows. In Sec.~\ref{sec:deformation},
we discuss the possible deformation of magnetars. The free and forced precession
dynamics of general triaxially-deformed magnetars are studied in
Sec.~\ref{sec:precession}. We give the phase modulations and timing residuals
due to precession in Sec.~\ref{sec:timing}. The modulations on polarized X-ray
and radio signals are investigated in Sec.~\ref{sec:polarization}. We give 
discussions and conclusions in Sec.~\ref{sec:discussion} and Sec.~\ref{sec:conclusion} respectively.

\section{Deformation of magnetars}
\label{sec:deformation}

To make the body precess, NSs must have some deformation misaligned with the
rotational bulge. The strains in the solid crust and the strong internal
magnetic fields are usually considered as potential causes of the deformation.
In the following, we consider the possible sources of deformation.
In the body frame, we can write the moment of
inertia tensor of a slowly rotating NS as~\citep{Jones:2000ud,Wasserman:2021dlh}
\begin{equation}
	I_{i j}=I_{0}\left[\delta_{i j}+\epsilon_{\mathrm{rot}}\left(\frac{1}{3} \delta_{i j}-
	\hat{\boldsymbol{\omega}}_{i} \hat{\boldsymbol{ \omega}}_{j}\right)+ M_{i j}
	\right]\,.
\end{equation}
Here $I_{0}$ is the spherical part of the moment of inertia. The rotational 
deformation is quadrupolar that is symmetric about the spin axis $\boldsymbol {\hat{\omega}}$.
We let $\epsilon_{\rm rot}$ denote the ellipticity sourced from the centrifugal force,
while the magnetic and elastic deformations are not necessarily axisymmetric. Thus, we use 
the symmetric and trace free (STF) tensor $M_{ij}$ to describe the deformations sourced from
the internal magnetic field, elasticity in the crust, or the combination of both.

For magnetars, the spin period $P$ is on the order of several seconds and
$\epsilon_{\rm rot}$ can be approximated as the rotational energy over the
gravitational energy
\begin{equation}
	\epsilon_{\rm rot} \approx \frac{\omega^{2} R^{3}}{G M} = 8.5 \times 10^{-9}P_{5}^{-2} R_{6}^{3} / M_{1.4}\,,
\end{equation}
where $\omega=2\pi/P$ is the spin angular frequency, $P_{5}$ is the spin period
in units of $5\,\rm s$, $R_{6}$ is the NS radius $R$ in units of
$10^{6}\,\rm cm$, and $M_{1.4}$ is the NS mass $M$ in units of
$1.4\,M_{\odot}$. The centrifugal deformation is of no importance for free
precession~\citep{Glampedakis:2010qm,Wasserman:2021dlh}, which can be understood as follows. The
angular momentum for a freely-precessing NS can be written as
\begin{align}
	L_{i}&=I_{0}\left[\left(1-\frac{2 \epsilon_{\mathrm{rot}}}{3}\right)\delta_{ij} \omega_{j}+M_{i j} \omega_{j}\right] 
	= I_{0}^{\prime}\left(\delta_{i j}+M_{i j}^{\prime}\right) \omega_{j}\,,
\end{align}
where Einstein summation is used, $I_{0}^{\prime}=I_{0}(1-2\epsilon_{\rm rot}/3)$, 
and $M_{ij}^{\prime}=M_{ij}/(1-2\epsilon_{\rm rot}/3)$. Since the rotational
ellipticity $\epsilon_{\rm rot}$ is quite small, we can absorb the rotational
bulges into the spherical part and approximately rewrite the moment of inertia
tensor as 
\begin{equation}
	I_{i j}\simeq I_{0}\left(\delta_{i j}+ M_{i j}\right)\,.
\end{equation}

We let $\boldsymbol{\hat{e}}_{1}$, $\boldsymbol{\hat{e}}_{2}$, and
$\boldsymbol{\hat{e}}_{3}$ denote the three unit eigenvectors along the
principal axes of moment of inertia tensor $I_{ij}$ with corresponding
eigenvalues $I_{1}\leq I_{2} \leq I_{3}$.
The angular velocity is
$\boldsymbol{\omega}=\omega_{1}\boldsymbol{\hat{e}}_{1}+
\omega_{2}\boldsymbol{\hat{e}}_{2}+\omega_{3}\boldsymbol{\hat{e}}_{3}$ and the
angular momentum is $\boldsymbol{L}=L_{1}\boldsymbol
{\hat{e}}_{1}+L_{2}\boldsymbol{\hat{e}}_{2}+L_{3}\boldsymbol{\hat{e}}_{3}$.
To describe the motion of the body, we define 
\begin{align}
	\epsilon &\equiv \frac{I_{3}-I_{1}}{I_{1}}, \nonumber\\
	\delta &\equiv \frac{I_{3}(I_{2}-I_{1})}{I_{1}(I_{3}-I_{2})}, 
	\nonumber \\
	\theta &\equiv \arccos\frac{L_{3}}{L}\,,
\end{align}
where $\epsilon$ is the ellipticity, $\delta$ measures the deviation from
axisymmetry, and $\theta$ is the wobble angle between $\boldsymbol
{\hat{e}}_{3}$ and $\boldsymbol{L}$.

Before investigating the dynamics of free precession, we first give an estimation of $\epsilon$.
The shear stresses of the crystallized solid crust can prevent a small fraction of 
the hydrostatic rotational bulge from aligning with the instantaneous spin axis.
We denote the ellipticity sourced from elastic deformation as $\epsilon_{\rm c}$.
The upper limit of $\epsilon_{\rm c}$ is approximated 
as~\citep{Ushomirsky:2000ax,Haskell:2006sv,Johnson-McDaniel:2012wbj,Gittins:2020cvx,Morales:2022wxs}
\begin{equation}
	\epsilon_{\mathrm{c}} ^{\max } \approx 10^{-6}\left(\frac{\sigma_{\mathrm{br}}}{10^{-1}}\right)\,,
\end{equation}
where the breaking strain $\sigma_{\rm br}\sim 0.1$ is taken from the molecular dynamics
simulations of crustal fracture in~\citet{Horowitz:2009ya}.  The actual value
of $\epsilon_{\rm c}$ depends on the evolution history of the star and is hard
to estimate. It may be much smaller than $\epsilon_{\mathrm{c}} ^{\max }$ since
plastic processes may relieve the strain in long time evolution. 

The magnetic fields inside the star create deformation because non-radial field
gradients can support non-radial matter-density gradients in hydromagnetic
equilibrium. However, the strength and geometry of the internal magnetic fields
are still very uncertain.  The magnetic ellipticity $\epsilon_{\rm B}$ is on the
order of the magnetic energy over the gravitational energy
\begin{equation}
	\epsilon_{\mathrm{B}}=\frac{\kappa H \bar{B} R^{4}}{M^{2}} \approx 1.93\times 10^{-6} 
	\kappa\,R_{6}^{4} M_{1.4}^{-2}H_{15}\bar{B}_{15}\,,
\end{equation}
which is a crude estimation but consistent with more rigorous calculations 
\citep[see
e.g.,][]{Haskell:2007bh,Akgun:2007ph,Lander:2009ib,Ciolfi:2010td,Mastrano:2013jaa}.
Here $\bar B$ is the volume average of the internal magnetic field, ${\bar
B}_{15}$ represents the magnetic field in units of $10^{15}\,\rm G$,
$H=\bar{B}$ for a normal conducting interior while $H\simeq 10^{15}\,\rm G$ if the core
sustains protons in the type~II superconducting 
state~\citep{Wasserman:2002ec,Cutler:2002np}. The parameter $\kappa$ can be
positive or negative depending on the relative strength between the poloidal and
toroidal components of the internal magnetic field. One can quickly notice that,
magnetars with large internal magnetic fields may cause large magnetic
deformation.

Most earlier studies have been devoted to axisymmetric magnetic field, regardless of whether the magnetic field is poloidal, toroidal or ``mixed''. The star is
deformed into a biaxial shape with the deformation axis along the dipole field.
We relax this axisymmetric assumption and study a more general case with
triaxial deformations and misalignment between the magnetic dipole and
deformation axes. On the one hand, the tilted poloidal-toroidal configuration is
more general and not physically
forbidden~\citep{Lasky:2013bpa,Wasserman:2021dlh}. For instance, \citet{Lasky:2013bpa}
obtained a tilted torus magnetic field from magnetohydrodynamic (MHD)
simulation, which is stable but the equilibrium state could not be specified
freely~\citep{Glampedakis2016}.  On the other hand, multipolar magnetic fields
may exist in the interior of magnetars.
\citet{Mastrano:2013jaa,Mastrano:2015rfa} found that the mixed odd and even
multipoles can create a deformation that is misaligned with the magnetic dipole
axis even if the magnetic field is axisymmetric. Moreover, the mixture of
elastic and magnetic deformations can produce a triaxial
shape~\citep{Wasserman:2002ec,Glampedakis:2010qm}.

The relative strength between the poloidal and toroidal fields is also not
clear. There is no stable mixed poloidal-toroidal field in NSs for a barotropic
normal fluid~\citep{Lander:2012iz}. According to MHD simulations,
\citet{Braithwaite:2008aw} argued that an axisymmetric field is stable in
stratified fluid if the poloidal field is much weaker than the toroidal field.
In this case, the deformation is prolate. \citet{Lander:2012zz} presented the
first self-consistent superconducting NS equilibria with poloidal and mixed
poloidal-toroidal fields. The poloidal component was dominant in all the
configurations that \citet{Lander:2012zz} studied.

According to the above discussions on the deformations and the structures of the
internal magnetic fields, we give following arguments and assumptions.
\begin{enumerate}
	\item Generally, the deformed NS is in a triaxial shape. The biaxial case is
	only a good approximation if the deformation is along a specific axis.
    \item The external magnetic axis is not necessarily aligned with any
    deformation axes. 
    \item The ellipticity can be positive or negative depending on the magnetic
    field geometry and possible elastic deformations.
    \item The precession of magnetically distorted NSs differs qualitatively
    from the elastic body precession although they have the same mathematical
    form.  There are slow and non-rigid internal motions in addition to the
    uniform rotation~\citep{Mestel1972,Lander:2016hwz}. Although these non-rigid
    motions are important for the evolution of the magnetic inclination angle,
    we ignore them in this work since they are higher-order effects. 
\end{enumerate}

\section{Dynamics of precession}
\label{sec:precession}

\subsection{Free precession}
\label{sec:free_precession}
 
The Euler equation of a freely-precessing body is~\citep{landau1960course}
\begin{equation}
	\label{eqn:free}
	\dot{\boldsymbol{L}}+\boldsymbol \omega \times \boldsymbol{L}=0\,,
\end{equation}
where the dot denotes the derivative with respect to time $t$. Eq.~(\ref{eqn:free}) can be solved
analytically in terms of Jacobian elliptic functions~\citep{landau1960course,Wasserman:2002ec,Akgun:2005nd,Gao:2020zcd,Wasserman:2021dlh}.
The angular momentum $\boldsymbol{L}$ and the kinetic energy $E$ are conserved
for free precession. Different branches of the solutions are determined by the 
relation between $L^{2}$ and $2EI_{2}$. One usually sets $\omega_{2}=0$ at the
initial time $t=0$. Thus, the solutions are also equivalently determined by the
parameter 
\begin{equation}
	m=\delta \tan^{2}\theta_{0}\,,
\end{equation}
with $\theta_{0}$ denoting the wobble angle $\theta$ at $t=0$.

\begin{figure}
    \centering
    \includegraphics[width=7.5cm]{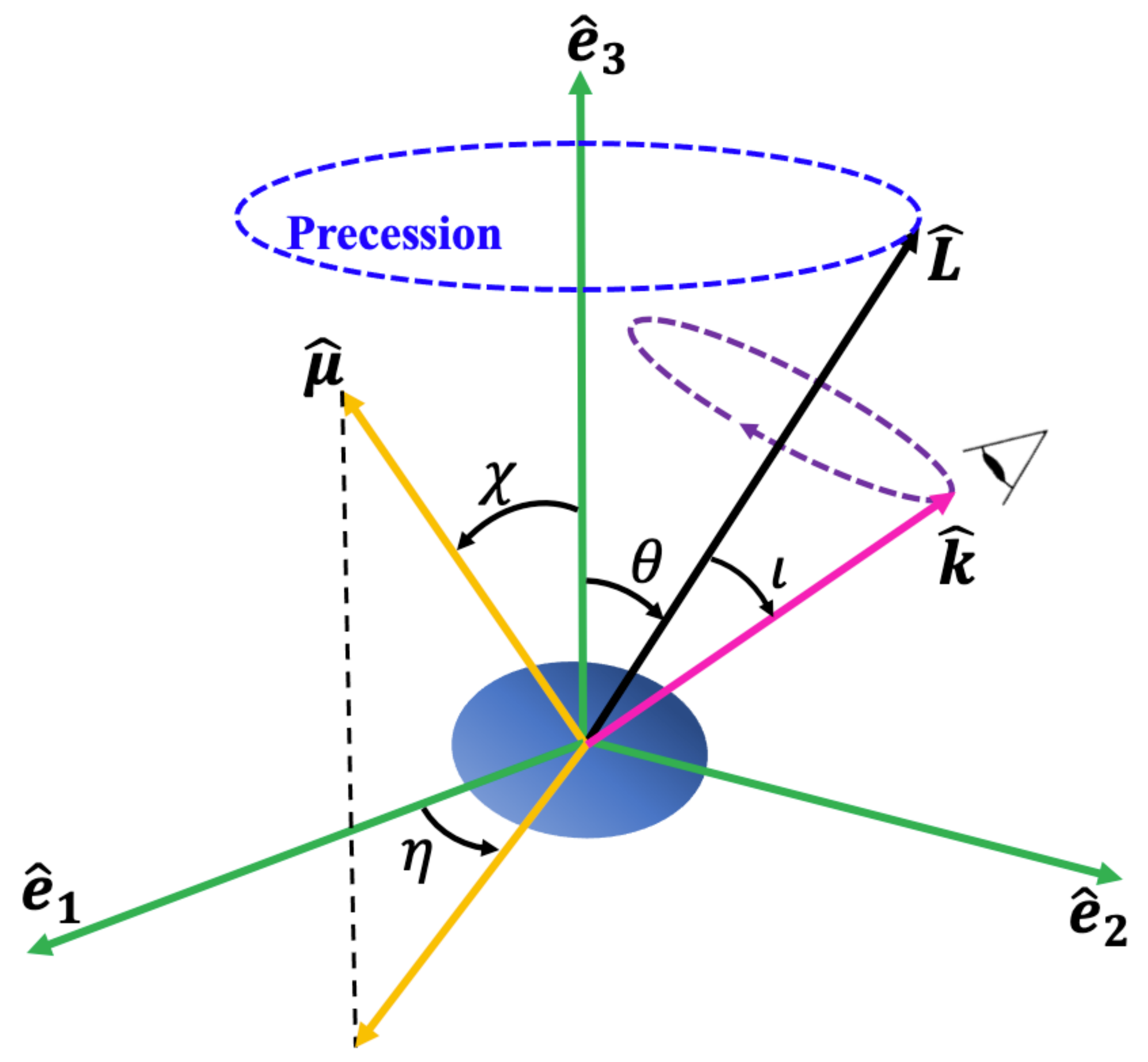}
    \caption{The geometry and the motion of a deformed NS in the corotating body
    frame. The NS precesses around $\hat{\boldsymbol{e}}_{3}$ with period $T$,
    which is called the {\it free precession period} of the NS. The wobble angle
    $\theta$ between $\hat{\boldsymbol{e}}_{3}$ and $\hat{\boldsymbol{L}}$
    nutates with a period of $T/2$. For the special biaxial case, the wobble
    angle $\theta$ is fixed. An observer views the NS in a direction
    $\hat{\boldsymbol{k}}$ fixed in the inertia frame but rotating around
    $\hat{\boldsymbol{L}}$ in the body frame with spin angular frequency
    $\boldsymbol{\omega}$ with an inclination angle $\iota$. The direction of
    the dipole moment $\hat{\boldsymbol{\mu}}$ is fixed in the body frame and is
    described by the polar angle $\chi$ and the azimuthal angle $\eta$.}
    \label{fig:body_frame}
\end{figure}

When $m<1$ ($L^{2}>2EI_{2}$), the precession is around
$\boldsymbol{\hat{e}}_{3}$ and the components of the unit angular momentum 
$\widehat {\boldsymbol{L}}\equiv \boldsymbol{L}/L$ are
\begin{align}
	&\hat{L}_{1}=\sin \theta_{0} \operatorname{cn}\left(\omega_{\rm p}t, m\right)\nonumber \,,\\
	&\hat{L}_{2}=\sin \theta_{0}\sqrt{1+\delta} \operatorname{sn}\left(\omega_{\rm p}t, m\right)\nonumber\,, \\
	&\hat{L}_{3}=\cos\theta_{0} \operatorname{dn}\left(\omega_{\rm p}t,m\right)\,,
	\label{eqn:oblate}
\end{align}
where $\operatorname{cn}$, $\operatorname{sn}$, and $\operatorname{dn}$ are Jacobi elliptic functions (see Appendix~\ref{append:A} for more details), and
\begin{align}
	\omega_{\mathrm{p}}=\frac{\epsilon L \cos \theta_{0}}{I_{3}\sqrt{1+\delta}}\,.
\end{align}
The time evolution of the angular frequencies in the body frame is periodic with
a period
\begin{equation}
	T=\frac{4I_{3}}{\epsilon L \cos\theta_{0}}\sqrt{1+\delta}\,K(m)\,,
\end{equation}
where $K(m)$ is the complete elliptic integral of the first kind. One can notice
that $2\pi/\omega_{\rm p}$ is not equal to the period $T$ because the Jacobi
elliptic functions are not  periodic in $2\pi$, but rather  periodic in $4K(m)$.
In Fig.~\ref{fig:body_frame}, we illustrate the geometry and the motion in the
corotating body frame. The angular momentum precesses around
$\hat{\boldsymbol{e}}_{3}$ with a period $T$. \revise{Following the definition in \citet{Cutler:2000bp},} we call $T$ the free precession
period of the deformed NS. The wobble angle $\theta$ nutates with a period
$T/2$. 

When $m=1$, the solution is unstable and the trajectories of the angular
momentum will decay exponentially to the intermediate axis 
$\boldsymbol{\hat{e}}_{2}$. \revise{The detailed solution can be found in \citet{landau1960course}.}
We omit this special case. 

When $m>1$, the precession is around $\boldsymbol{\hat{e}}_{1}$ and the
solutions are given in~\citet{Akgun:2005nd} and \citet{Zanazzi:2015ida}.  We
want to retain the definition of $\theta$ as the angle between
$\hat{\boldsymbol{L}}$ and $\boldsymbol{\hat{e}}_{3}$, which is helpful to map
the latter calculations to the $m<1$ case directly. So, we make a redefinition
of the basis vectors, 
\begin{align}
	\boldsymbol{\hat{e}}_{1} &=\boldsymbol{\hat{e}}_{3}\,,\nonumber\\
	\boldsymbol{\hat{e}}_{2} &=-\boldsymbol{\hat{e}}_{2}\,,
	\nonumber\\
	\boldsymbol{\hat{e}}_{3} &=\boldsymbol{\hat{e}}_{1}\,.
\end{align}
Then the solutions of $\hat{\boldsymbol{L}}$ can be represented in an identical form
to the $m<1$ case, except for $I_{1}\geq I_{2} \geq I_{3}$. Note that $\epsilon<0$ and 
the precession direction is opposite to the $m<1$ case.

A biaxial body corresponds to the special cases of oblate deformation
$(\delta=0, \epsilon>0)$ or prolate deformation $(\delta=0, \epsilon<0)$.  
Eq.~(\ref{eqn:oblate}) degenerates into the form 
\begin{align}
	\hat{L}_{1}&=\sin \theta_{0} \cos(\omega_{\rm p}t) \nonumber\,,\\
	\hat{L}_{2}&=\sin \theta_{0}  \sin(\omega_{\rm p}t) \nonumber\,, \\
	\hat{L}_{3}&=\cos\theta_{0} \,,
	\label{eqn:axisymmetry}
\end{align}
where 
\begin{equation}
	\omega_{\rm p}=\epsilon L \cos \theta_{0}/I_{3}\,,
\end{equation}
becomes the precession frequency, and the precession period is 
\begin{equation}
	T=\frac{2\pi}{\lvert\omega_{\rm p}\rvert}=\frac{2\pi I_{3}}{\lvert\epsilon \rvert L \cos \theta_{0}}\,.
\end{equation}

In later examples, we only study the $m<1$ branch, the cases for $m>1$ can be
easily obtained by redefining $I_{1}\geq I_{2} \geq I_{3}$.
Although the precession period $T$ is different for distinct NS geometries, we
can roughly estimate the timescale of the precession as 
\begin{equation}
	\label{eqn:precession_timescale}
	\tau_{\rm p}=\frac{P}{\epsilon}=1.58P_{5}\epsilon_{7}^{-1}\,\rm yr\,,
\end{equation}
where $\epsilon_{7}=\epsilon/10^{-7}$. The angular velocity in the body frame is
\begin{equation}
	\boldsymbol{\omega}=L\left(\frac{\hat{\boldsymbol{L}}_{1}}{I_{1}}+\frac{\hat{\boldsymbol{L}}_{2}}{I_{2}}+
	\frac{\hat{\boldsymbol{L}}_{3}}{I_{3}}\right)=\frac{L}{I_{3}}\left(\frac{I_{3}\hat{\boldsymbol{L}}_{1}}{I_{1}}+
	\frac{I_{3}\hat{\boldsymbol{L}}_{2}}{I_{2}}+\hat{\boldsymbol{L}}_{3}\right)\,.
\end{equation}
It is obvious that the angle $\theta^{\prime}$ between the angular momentum $\boldsymbol{L}$ and
the angular velocity $\boldsymbol{\omega}$ is on the order of
$\theta^{\prime}\sim \epsilon \theta \ll 1$.  We can approximate $\boldsymbol{L}
\parallel \boldsymbol{\omega}$ to the zeroth order of $\epsilon$ when we
evaluate the geometry of the NS. We denote the unit vector of the angular
frequency as $\hat{\boldsymbol{\omega}}\equiv \boldsymbol{\omega}/\omega_{0}$,
where $\omega_{0}$ is the magnitude of the angular frequency at $t=0$. For
simplicity, we also take the external magnetic field as a dipole field. The
dipole moment, $\boldsymbol{\mu}=\mu \hat{\boldsymbol{\mu}}$, is fixed in the
body frame and
\begin{equation}
	\label{eqn:dipole_body}
	\hat{\boldsymbol{\mu}}=\hat{\mu}_{1}\hat{\boldsymbol{e}}_{1}+\hat{\mu}_{2}\hat{\boldsymbol{e}}_{2}
	+\hat{\mu}_{3}\hat{\boldsymbol{e}}_{3}=\left(\sin\chi\cos\eta\,, \sin\chi\sin\eta\,,\cos\chi\right)\,.
\end{equation}
Note that the angle $\eta$ is not necessarily zero in the general triaxial case.
Only in the biaxial case can one choose $\eta=0$ due to the axisymmetry of the
NS.

\subsection{Forced precession}
\label{sec:forced_precession}

\begin{figure}
    \centering
    \includegraphics[width=8.5cm]{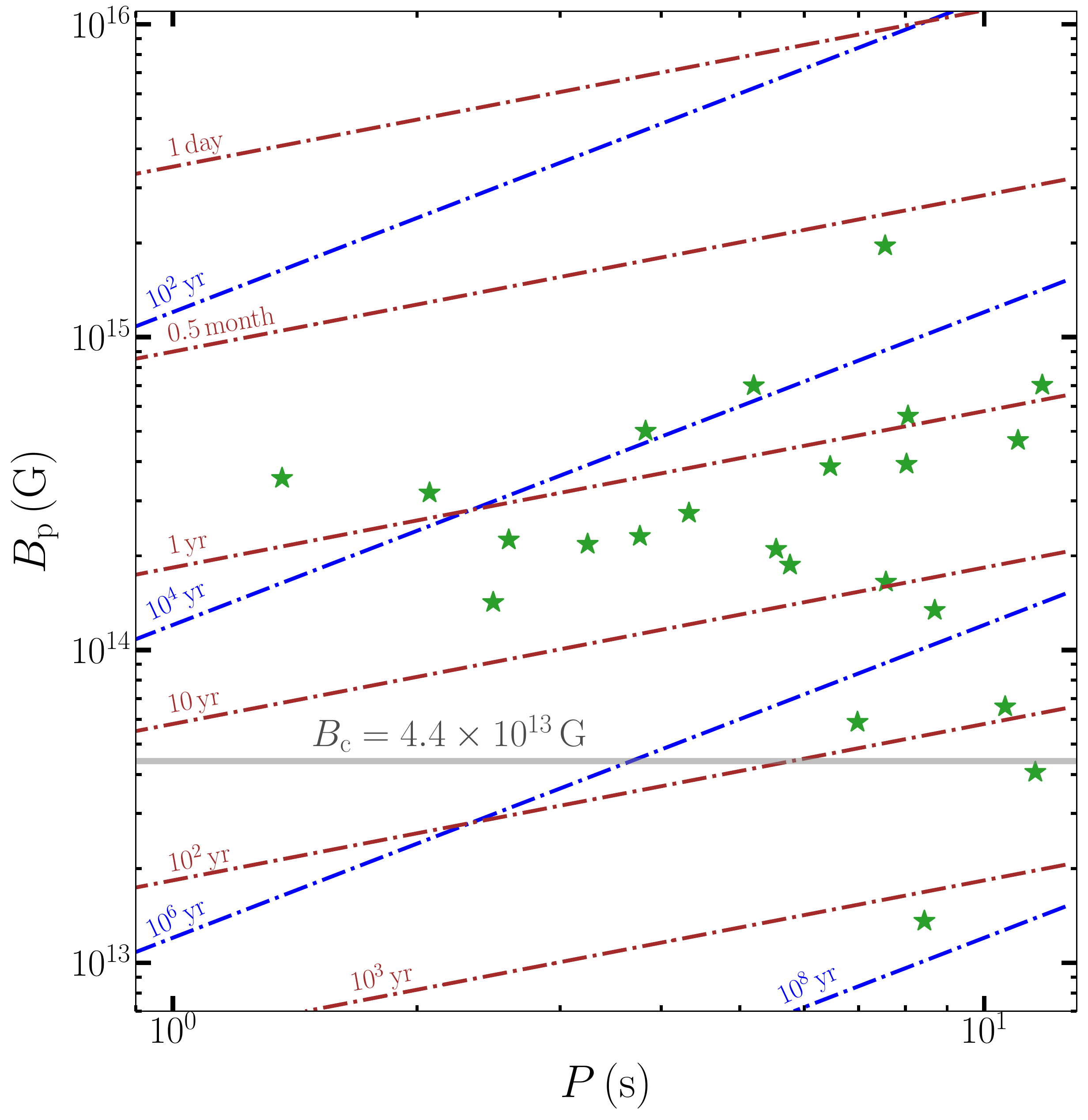}
    \caption{The relation between the spin period $P$ and the magnetic field at
    the magnetic pole $B_{\rm p}$ for magnetars with measured period and period
    derivative (green). Lines of constant $\tau_{\rm rad}$ (blue) and constant
    $\tau_{\rm m}$ (brown) are also illustrated. The horizontal gray line
    represents the Schwinger limit of the magnetic field, $B_{\rm c}=4.4\times
    10^{13}\,\rm G$.}
    \label{fig:timescale}
\end{figure}

For magnetars, large magnetic fields also indicate large electromagnetic
torques. Generally, rotating NSs endowed with external magnetic fields have two
kinds of electromagnetic torques acting on it. The first one is the far-field
torque (the so-called spin-down torque), which originates from the fact that the
electromagnetic emission carries away angular
momentum~\citep{Deutsch,Davis1970}. For a dipole field, we express the far-field torque
$\boldsymbol{N}_{\rm rad}$ as 
\begin{equation}
	\label{eqn:spindown}
	\boldsymbol{N}_{\rm rad}=\frac{k_{1} \mu^{2} \omega^{3}}{c^{3}}\left[(\hat{\boldsymbol{\omega}} \cdot \hat{\boldsymbol{\mu}}) 
	\hat{\boldsymbol{\mu}}-k_{2}\hat{\boldsymbol{\omega}} \right]\,,
\end{equation}
where $k_{1}$ and $k_{2}$ are numerical constants on the order of unity. 

For the simplest vacuum dipole, $k_{1}=2/3$,
$k_{2}=1$~\citep{Deutsch,Davis1970}, and the rotational energy dissipates at a
rate 
\begin{equation}
	\boldsymbol{N}_{\rm rad}\cdot \boldsymbol{\boldsymbol{\omega}}=-\frac{2\mu^{2}\omega^{4}\sin^{2}\alpha}{3c^{3}}\,,
\end{equation}
where $\alpha$ is the magnetic inclination angle between
$\hat{\boldsymbol{\omega}}$ and $\hat{\boldsymbol{\mu}}$.  There is no
dissipation for the vacuum dipole case when the angular velocity and the dipole
moment are aligned, namely $\sin\alpha=0$. 

Although the vacuum magnetosphere torque predicts the spin-down rate somehow
close to the observed values for pulsars, the magnetosphere is filled with
plasma in reality. The charges and currents in the plasma inevitably modify the
structure of magnetosphere. There are no analytical expressions for the far-field
torque for a plasma-filled magnetosphere. \citet{Li:2011zh} and \citet{Philippov:2013aha} analyzed the
results of force-free MHD simulations and found that the plasma-filled torque
can be approximately parameterized by taking $k_{1}\simeq 1$ and $k_{2}\simeq 2$ if 
the weak dependence on $R/R_{\rm LC}$ is ignored, with $R_{\rm LC}$ being the radius of the light cylinder.
This parametrization of the far-field torque was also applied to study the precession of pulsars~\citep{Arzamasskiy:2015lza}.
In this case, the rotational energy dissipates at a rate
\begin{equation}
	\boldsymbol{N}_{\rm rad}\cdot \boldsymbol{\boldsymbol{\omega}}=-\frac{\mu^{2}\omega^{4}}{c^{3}} (1+\sin^{2}\alpha)\,.
\end{equation}
The energy still can be dissipated when $\alpha=0$ compared to the vacuum case.
The form of the plasma-filled torque is equivalent to adding a parallel component compared 
to the vacuum case. 

The far-field torque not only dissipates the rotational energy but changes the
geometry of the star, such as the wobble angle and the magnetic inclination
angle. We define the spin-down timescale induced by the far-field torque as 
\begin{equation}
	\tau_{\rm rad} = \frac{3c^{3}I_{0}}{2\mu^{2}\omega^{2}}=3.61 \times 10^{5} M_{1.4}P_{5}^{2}B_{14}^{-2}\,\rm yr\,.
\end{equation}

The second kind of the electromagnetic torque is the near-field torque, which
arises from the inertia of the external magnetic 
field~\citep{Davis1970,Good1985,Melatos1997}. The near-field torque is denoted
by $\boldsymbol{N}_{\rm m}$ and can be expressed as 
\begin{equation}
	\label{eqn:near_field}
	\boldsymbol{N}_{\rm m}=\frac{k_{3}\omega^{2}\mu^{2}}{R c^{2}}(\hat{\boldsymbol{ \omega}} \cdot 
	\hat{\boldsymbol{\mu}})(\hat{\boldsymbol{\omega }}\times \hat {\boldsymbol{\mu}})\,,
\end{equation}
where the external magnetic field is assumed to be a dipole field.  Using
different methods, many authors have obtained slightly different values of
$k_{3}$~\citep{Goldreich:1970ads,Good1985,Melatos:2000qt,Beskin:2013zqa,Zanazzi:2015ida}.
Here, we adopt the value $k_{3}=3/5$, which is consistent
with~\citet{Melatos1997} and~\citet{Zanazzi:2015ida}. This value can be obtained
by assuming a uniform internal magnetic field $\boldsymbol{B}_{\rm p}$ rotating
rigidly around the spin axis, and the electric field given by
$\boldsymbol{E}=-(\boldsymbol{v} / c) \times \boldsymbol{B}_{\rm p}$ for a perfectly
conducting fluid. Although an internal electromagnetic field is assumed, the
near-field torque in Eq.~(\ref{eqn:near_field}) only depends on the exterior
electromagnetic field of the NS~\citep{Beskin:2014mva,Zanazzi:2015ida}. 

The near-field torque is perpendicular to $\boldsymbol{\omega}$ and scales as
$\omega^{2}$. It does not dissipate energy or angular momentum but affects the
spin and the wobble angle of the precessing NS in a timescale of 
\begin{equation}
	\tau_{\rm m}=\frac{5RI_{0}c^{2}}{3\omega\mu^{2}}=16.8 M_{1.4}R_{6}P_{5}B_{14}^{-2}\,\rm yr\,.
\end{equation}

In Fig.~\ref{fig:timescale}, we plot the relation between the dipole magnetic
field at the magnetic pole $B_{\rm p}$ and the rotation period $P$ for magnetars
with measured period and period
derivative~\citep{Olausen:2013bpa}.\footnote{\url{http://www.physics.mcgill.ca/~pulsar/magnetar/main.html}}
We also plot the contour
lines for $\tau_{\rm m}$ and $\tau_{\rm rad}$. For typical magnetars with
$B_{\rm p}\sim 10^{14}$--$10^{15}\,\rm G$, $\tau_{\rm m}$ is on the order of
$0.1$--$10\,\rm yr$ and $\tau_{\rm rad}$ is on the order of
$10^{3}$--$10^{5}\,\rm yr$. It is very interesting that $\tau_{\rm m}$ and the
free precession timescale $\tau_{\rm p}$ could be comparable in some cases, and if so
the free precession solution will be affected substantially.
\citet{Melatos1997,Melatos:2000qt} studied this effect and gave detailed
numerical solutions for different NS geometries. In our work, we adopt an
analytical method developed by~\citet{Glampedakis:2010qm} and
\citet{Zanazzi:2015ida} to study the precession dynamics under the near-field
torque. Since $\tau_{\rm rad}$ is much larger than $\tau_{\rm m}$ and $\tau_{\rm
p}$, we use a perturbative method to study the forced precession under the
far-field torque following~\citet{Goldreich:1970ads}, \citet{Link:2001zr},
\citet{Wasserman:2002ec}, and \citet{Wasserman:2021dlh}. 

\subsubsection{Near-field torque}
\label{sec:dynamics_near}

Under the near-field torque, the Euler equation is
\begin{equation}
	\label{eqn:near1}
	\dot{\boldsymbol{L}}+\boldsymbol \omega \times \boldsymbol{L}= \frac{3\omega^{2}\mu^{2}}{5R c^{2}}(\hat{\boldsymbol{ \omega}} 
	\cdot \hat{\boldsymbol{\mu}})(\hat{\boldsymbol{\omega }}\times \hat {\boldsymbol{\mu}})\,.
\end{equation}
The near-field torque arises from the inertia of the electromagnetic field,
which can actually be absorbed into the moment of inertia tensor
$\boldsymbol{I}$ of the
star~\citep{Melatos:2000qt,Glampedakis:2010qm,Zanazzi:2015ida}. 
Eq.~(\ref{eqn:near1}) can be written as 
\begin{equation}
	\dot{\boldsymbol{L}}+\boldsymbol \omega \times (\boldsymbol{L}+\boldsymbol{\omega}\cdot\boldsymbol{M})=0\,,
\end{equation}
by introducing the effective deformation tensor
\begin{equation}
	\boldsymbol{M}=-I_{0} \epsilon_{\rm m}(\hat{\boldsymbol{\mu}} \otimes \hat{\boldsymbol{\mu}})\,,
\end{equation}
Here, $\epsilon_{\rm m}$ is the effective ellipticity induced by the external
magnetic field and 
\begin{equation}
	\epsilon_{\rm m}=\frac{3\mu^{2}}{5I_{0}Rc^{2}}=1.5\times 10^{-9}M_{1.4}^{-1}B_{14}^2R_{6}^{3}\,.
\end{equation}
Since $\epsilon_{\rm m}$ is quite small, we can introduce an effective moment of
inertia tensor $\boldsymbol{I}_{\rm eff}=\boldsymbol{I}+\boldsymbol{M}$ and
write the Euler equations as~\citep{Zanazzi:2015ida}
\begin{equation}
	\dot{\boldsymbol{L}}_{\rm eff}+\boldsymbol \omega \times \boldsymbol{L}_{\rm eff}=0\,,
\end{equation}
with the effective angular momentum $\boldsymbol{L}_{\rm
eff}=\boldsymbol{I}_{\rm eff}\cdot \boldsymbol{\omega}$.  Thus, the forced
precession under the near-field torque is transformed into free precession by
introducing a new prolate deformation along the magnetic dipole axis with an
ellipticity $\epsilon_{\rm m}$. In principle, one can solve the
forced-precession problem in Eq.~(\ref{eqn:near1}) numerically, but the
transformation used here gives analytical solutions and more insight into this problem.
Therefore, we give the analytical solutions of the free precession in the
effective principal frame. In practice, one just needs to substitute all the
quantities in Sec.~\ref{sec:free_precession} into corresponding effective ones.

In Sec.~\ref{sec:free_precession}, we assumed $\omega_{2}=0$ at $t=0$ for
simplicity. The phase of the precession is just $\omega_{\rm p}t$.  For
consistency, one must be cautious of the initial phase when calculating the
effective problem. For a general triaxial star, the magnetic dipole moment does
not necessarily lie in the $\hat{\boldsymbol{e}}_{1}$-$\hat{\boldsymbol{e}}_{3}$
plane ($\eta \neq 0$). Thus, the effective deformation caused by the near-field
torque makes $\omega_{{\rm eff}, 2}\neq 0$ at the initial time. So, the
precession phase of the solutions should be $\omega_{\rm p, eff}\,t+\psi_{0}$,
where 
\begin{equation}
	\omega_{\mathrm{p, eff}}=\frac{\epsilon_{\rm eff} L_{\rm eff} \cos \theta_{0,\rm eff}}{I_{3,\rm eff} \sqrt{1+\delta_{\rm eff}}}\,,
\end{equation}
and the initial phase
\begin{equation}
	\psi_{0}=-\arcsin \sqrt{L_{1,\rm eff}^2 + \frac{L_{2,\rm eff}^2}{1+\delta_{\rm eff}}}.
\end{equation}
Only in the triaxial case with $\eta=0$ and the special biaxial case, one can
take $\psi_{0}=0$. Here, we denote the basis vectors, the eigenvalues of the
moment of inertia tensor, the components of the angular velocity, the related
geometric parameters in the effective principal frame with a lower index ``eff''
based on the quantities in the original principal frame. Generally, we find that 
the effects of the near-field torque can be ignored only if $\epsilon_{\rm m}\lesssim 0.1\epsilon$.

\subsubsection{Far-field torque}
\label{sec:dynamics_far}

\def\arraystretch{1.3}
\begin{table*}
	\caption{The intrinsic and effective parameters for forced precession shown in Fig.~\ref{fig:spindown1}-\ref{fig:spindown2}.}
	\centering 
	\begin{tabular}{lccccccccccccc}
		\toprule
		 Case&\multicolumn{7}{c}{Intrinsic parameters}&\multicolumn{6}{c}{Effective parameters} \\
		 \cmidrule(lr){2-8}\cmidrule(lr){9-14}
		 &
		$P_{0}\,(\rm s)$&
		$B\,(\rm G)$ & 
		$\epsilon$ & 
		$\delta$ & 
		$\theta_{0}\, (\degr)$ & 
		$\chi\, (\degr)$ & 
		$\eta \, (\degr) $&
		$\epsilon_{\rm eff}$ &
		$\delta_{\rm eff}$ & 
		$\theta_{\rm eff, 0}\, (\degr)$ & 
		$\chi_{\rm eff}\, (\degr) $ & 
		$\eta_{\rm eff} \, (\degr) $ & 
		$T_{\rm eff}\,(\rm yr)$ \\ 
		\hline
		\uppercase\expandafter{\romannumeral1} & 5 & $5\times 10^{14}$ & $10^{-7}$ & 0 & 10&  45 & 0 & $1.07\times 10^{-7}$& 0.261& 20.3&55.3 & 0 & 1.79 \\
	
		\uppercase\expandafter{\romannumeral2} & 5 & $5\times 10^{14}$ & $10^{-7}$ & 1 & 10&  45 & 45 & $9.99\times 10^{-8}$& 1.15& 20.5&60.2 & 36.1 & 2.59 \\
		
		\uppercase\expandafter{\romannumeral3} & 5 & $ 10^{14}$ & $10^{-7}$ & 0 & 10&  45 & 0 & $1.00\times 10^{-7}$& 0.007& 10.4& 45.4 & 0 & 1.62 \\
		
		\uppercase\expandafter{\romannumeral4} & 5 & $ 10^{14}$ & $10^{-7}$ & 1 & 10&  45 & 45 & $9.96\times 10^{-8}$& 1.01& 10.3&45.6 & 45.6 & 2.31 \\
		\bottomrule
	\end{tabular}
	\label{tab:first}
\end{table*}


\begin{figure}
    \centering
    \includegraphics[width=8cm]{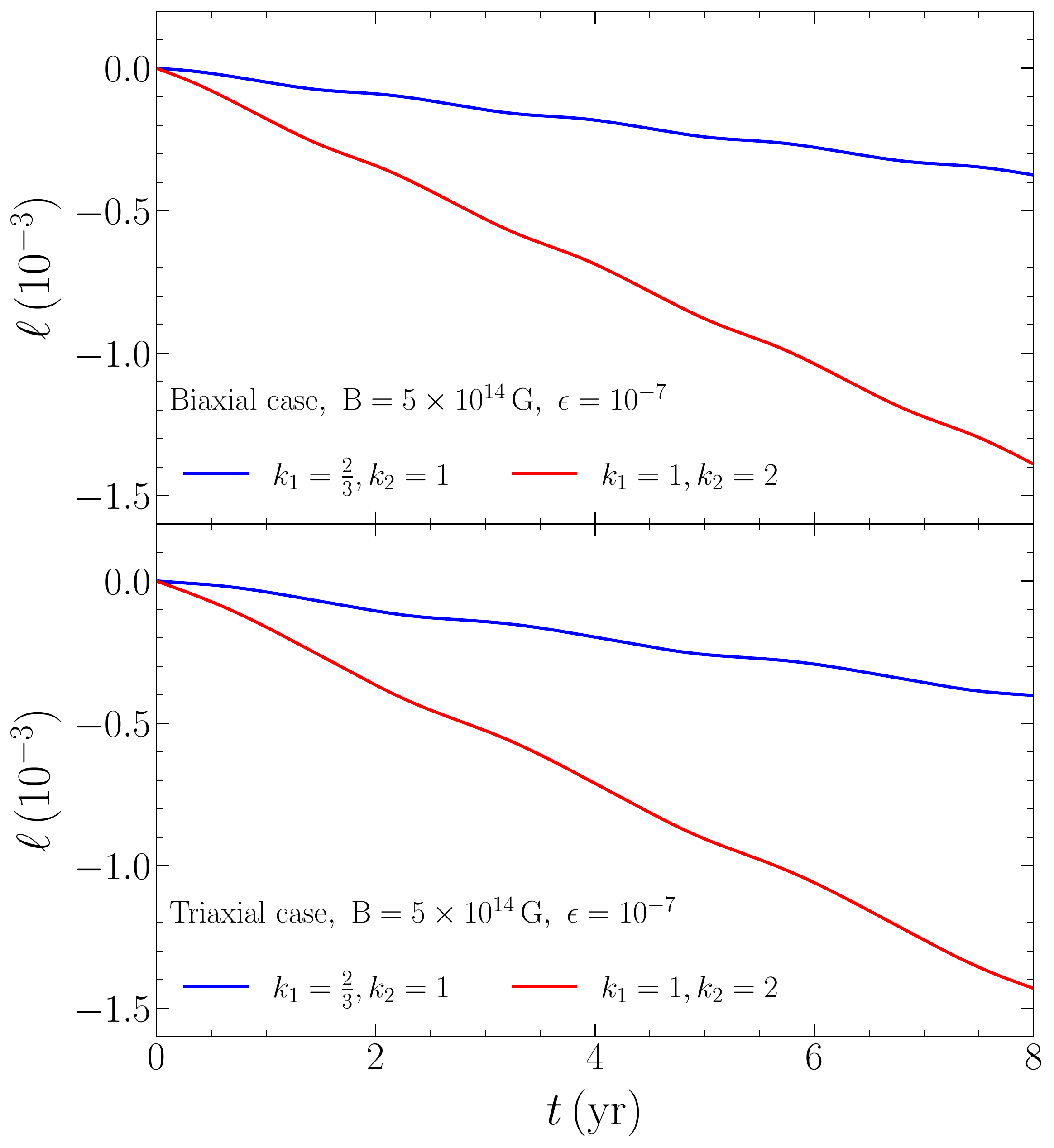}
    \caption{The fractional change of the angular frequency due to spin down for
    the biaxial case (upper) and the triaxial case (lower). For comparison, both of
    the vacuum torque ($k_{1}=2/3, k_{2}=1$) and the plasma-filled torque
    ($k_{1}=1, k_{2}=2$) are illustrated. The parameters for the biaxial and triaxial 
	cases are shown in Case \uppercase\expandafter{\romannumeral1} and Case 
	\uppercase\expandafter{\romannumeral2} of Table~{\ref{tab:first}} respectively.}
    \label{fig:spindown1}
\end{figure}

\begin{figure}
    \centering
    \includegraphics[width=8cm]{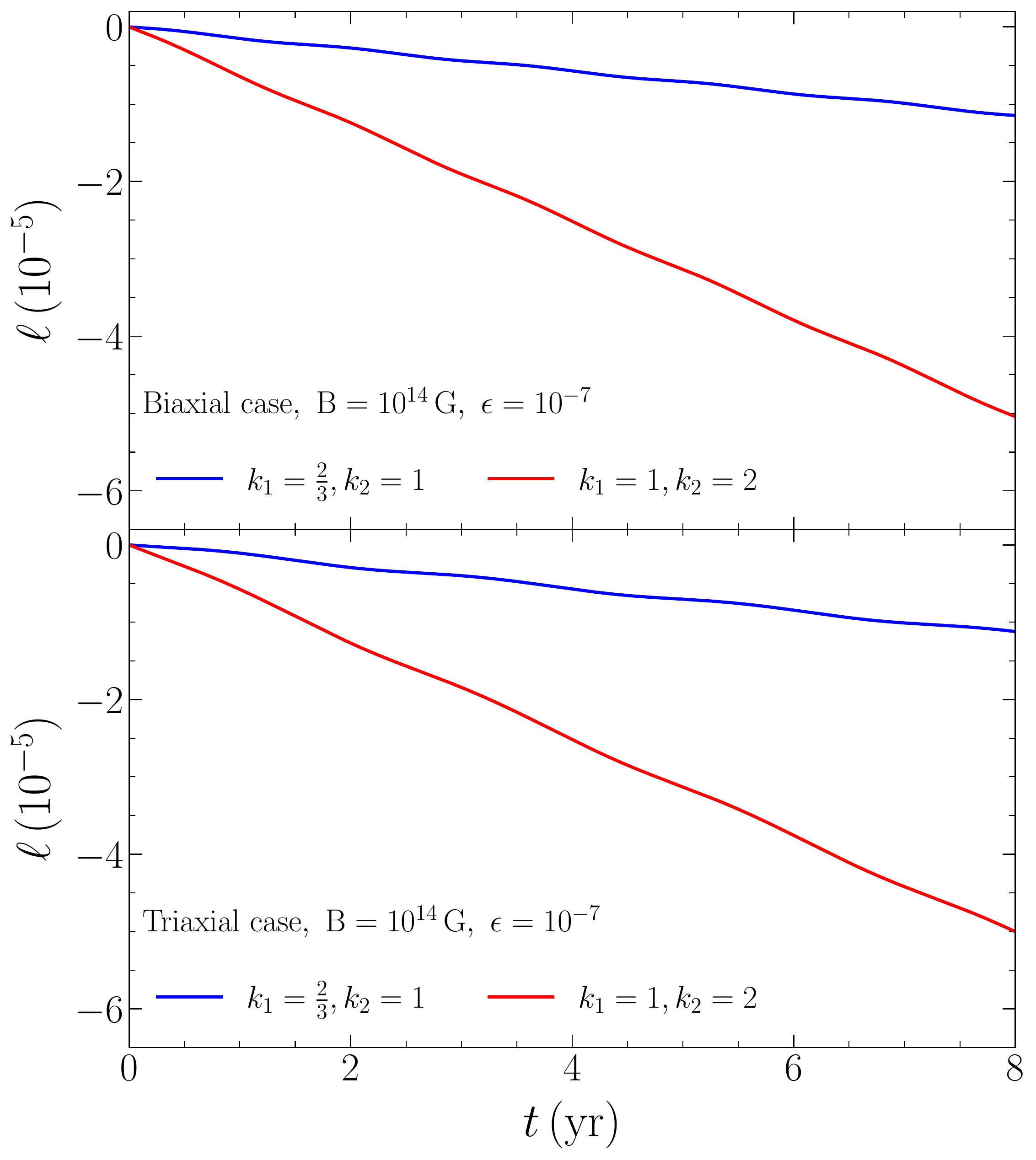}
    \caption{Same as Fig.~\ref{fig:spindown1} but for a deformed magnetar with
    $B=10^{14}\,\rm G$. The parameters for the biaxial and triaxial 
	cases are shown in Case \uppercase\expandafter{\romannumeral3} and Case 
	\uppercase\expandafter{\romannumeral4} of Table~{\ref{tab:first}} respectively.}
    \label{fig:spindown2}
\end{figure}

After absorbing the near-field torque into the effective moment of inertia
tensor, the Euler equation under the far-field torque can be written as 
\begin{equation}
	\label{eqn:euler_far_field}
	\dot{\boldsymbol{L}}_{\rm eff}+\boldsymbol \omega \times \boldsymbol{L}_{\rm eff}=\boldsymbol{N}_{\rm rad}=\frac{k_{1} 
	\mu^{2} \omega^{3}}{c^{3}}\left[(\hat{\boldsymbol{\omega}} \cdot \hat{\boldsymbol{\mu}}) \hat{\boldsymbol{\mu}}-k_{2}\hat{\boldsymbol{\omega}} \right]\,.
\end{equation}
For simplicity, we omit the ``eff'' notation in later equations and only give
``effective'' parameters in specific examples.  The far-field torque can be
decomposed into parallel and perpendicular components with respect to the
angular momentum
\begin{align}
	\boldsymbol{N}_{\rm rad}^{\parallel}=&\frac{k_{1} \mu^{2} \omega^{3}}{c^{3}}\left[(\hat{\boldsymbol{\omega}}\cdot 
	\hat{\boldsymbol{\mu}})(\hat{\boldsymbol{L}}\cdot \hat{\boldsymbol{\mu}}) -k_{2}(\hat{\boldsymbol{L}}\cdot 
	\hat{\boldsymbol{\omega}}) \right]\hat{\boldsymbol{L}}\,,\\
	\boldsymbol{N}_{\rm rad}^{\perp}=&\boldsymbol{N}_{\rm rad}-\boldsymbol{N}_{\rm rad}^{\parallel}=\boldsymbol{N}_{\rm rad}-
	(\hat{\boldsymbol{L}}\cdot \hat{\boldsymbol{N}})\hat{\boldsymbol{L}}\,.
\end{align}
Taking the dot product between Eq.~(\ref{eqn:euler_far_field}) and
$\hat{\boldsymbol{L}}$, we obtain 
\begin{equation}
	\label{eqn:euler_f1}
	 \dot{L}=\boldsymbol{N}_{\rm rad}^{\parallel}\cdot \hat{\boldsymbol{L}}\simeq 
	 \frac{3k_{1} I_{0}\omega}{2\tau_{\rm rad}}\left(\cos^{2}\alpha -k_{2} \right)\,,
\end{equation}
where $\hat{\boldsymbol{L}}$ and $\hat{\boldsymbol{\omega}}$ have been
approximated as the same direction on the right-hand side. 
Eq.~(\ref{eqn:euler_f1}) determines the magnitude of the angular momentum. The
angle $\alpha$ oscillates during the precession, which produces variability in
the spin-down rate. The perpendicular Euler equation is  
\begin{equation}
	\label{eqn:euler_f2}
	L\dot{\hat{\boldsymbol{L}}}+ \omega L(\hat{\boldsymbol{\omega}}\times \hat{\boldsymbol{L}})=\boldsymbol{N}_{\rm rad}^{\perp}\,,
\end{equation} 
which determines the direction of the angular momentum. The second term on the
left-hand side arises from the precession of the angular momentum around
$\hat{\boldsymbol{e}}_{3}$ in the body frame. The right-hand side is the
term that originates from the far-field torque, which contributes to the secular
change of $\theta$ and $\alpha$.

In this work, we concentrate on the spin evolution on the precession timescale.
According to Eq.~(\ref{eqn:euler_f2}), the angles $\theta$ and $\alpha$ change
secularly on the order of $\sim \tau_{\rm p}/\tau_{\rm rad}\ll 1$ under the
far-field torque. Thus, we can neglect the secular variation of $\alpha$ when
calculating the change of the angular momentum with Eq.~(\ref{eqn:euler_f1}).

To describe the spin evolution, we introduce 
\begin{equation}
	\omega(t)=\omega_{0}(1+\ell(t))\,,
\end{equation}
where $\omega_{0}$ is the angular frequency at the initial time $t=0$, and
$\ell$ is the fractional change of the angular frequency due to spin down. Since
$\tau_{\rm p}\ll \tau_{\rm rad}$ and the spin-down rate is quite small, we can
set $\omega=\omega_{0}$ on the right-hand side of Eq.~(\ref{eqn:euler_f1}) and
write the solution of $\ell$ as
\begin{align}
	\label{eqn:fractional}
	\ell= -\frac{3k_{1}}{2\tau_{\rm rad}}\left(k_{2}t-\int_{0}^{t}\cos^{2}\alpha \dd t\right) \,.
\end{align}
By introducing $\tau=\omega_{\rm p}t + \psi_{0}$, the magnetic inclination angle
$\alpha$ satisfies  
\begin{equation}
	\label{eqn:cosalpha}
	\cos\alpha = \hat{\mu}_{1}\sin\theta_{0}\operatorname{cn}\tau 
	+\hat{\mu}_{2}\sin\theta_{0}(1+\delta)^{1/2}\operatorname{sn}\tau
	+  \hat{\mu}_{3}\cos\theta_{0}\operatorname{dn}\tau\,,
\end{equation}
for the case of $m<1$, where the modulus $m$ has been omitted in the expressions
of Jacobi elliptic functions. Substituting Eq.~(\ref{eqn:int_alpha}) into
Eq.~(\ref{eqn:fractional}), one gets the fractional change of the angular 
frequency for different NS geometries. 

In Fig.~\ref{fig:spindown1}, we show the fractional change of the angular
frequency due to spin-down for both biaxial and triaxial cases with $B_{\rm
p}=5\times 10^{14}\,\rm G$ and $\epsilon=10^{-7}$. The spin-down rate oscillates
since $\alpha$ varies with the precession.  The ellipticity induced by the
near-field torque is $\epsilon_{\rm m}=3.75\times10^{-8}=0.375 \epsilon$. Thus,
the near-field torque affects the precession substantially. The initial wobble
angle for the motion is amplified from $10^{\circ}$ to about $20^{\circ}$, which
leads to large variations of the spin-down rate. We also plot the spin-down for
both of the vacuum torque and the plasma-filled torque. The angular velocity
decreases faster for the plasma-filled torque because the radiation power is
stronger. From Eq.~(\ref{eqn:fractional}), we also notice that the second 
term on the right-hand side only depends on the coefficient $k_{1}$.

We also give the case with $B_{\rm p}= 10^{14}\,\rm G$ and $\epsilon=10^{-7}$
in Fig.~\ref{fig:spindown2}. The ellipticity induced by the near-field torque is
$\epsilon_{\rm m}=1.5\times10^{-9}=0.015\epsilon$, which is negligible. The
effective wobble angle is nearly the same as the free precession case and the
variation of the spin-down rate is much smaller compared to the cases in
Fig.~\ref{fig:spindown1}. If $\epsilon \gtrsim 10^{-6}$ for the case with
$B_{\rm p}=5\times 10^{14}\,\rm G$, the effects of the near-field torque can be
also neglected.

For the biaxial case with $\epsilon_{\rm m}\ll \epsilon$ or the effective
biaxial case, the parameters $\delta$ and $\eta$ can be set to zero. The angle
$\alpha$ satisfies 
\begin{equation}
	\label{eqn:biaxial_alpha}
	\cos\alpha=\sin\chi\sin\theta_{0}\cos\omega_{\rm p}t +\cos\chi\cos\theta_{0}\,,
\end{equation}
and the integration of $\cos\alpha$ simplifies into
\begin{align}
	\label{eqn:biaxial_spindwon}
	\int_{0}^{t}\cos^{2}\alpha \dd t =&\frac{1}{4\omega_{\rm p}}\left[\left(\mu_{1}^{2}+2 \mu_{3}^{2}\right) \omega_{\rm p}t-
	\left(\mu_{1}^{2}-2 \mu_{3}^{2}\right) \omega_{\rm p}t\cos 2 \theta_{0} \right.\nonumber \\
	&\left.+4 \mu_{1} \mu_{3} \sin 2 \theta_{0} \sin \omega_{\rm p}t+\mu_{1}^{2} \sin ^{2} \theta_{0} \sin 2 \omega_{\rm p}t\right]\,.
\end{align}
where the initial phase $\psi_{0}=0$. The biaxial case in
Fig.~\ref{fig:spindown2} can be obtained by taking
Eq.~(\ref{eqn:biaxial_spindwon}) into Eq.~(\ref{eqn:fractional}) because
$\epsilon_{\rm m}\ll \epsilon$.

Our studies are similar to~\citet{Melatos1997}, but we give analytical
solutions and consider different models of the far-field torques. Compared
to~\citet{Akgun:2005nd} and \citet{Wasserman:2021dlh}, we have considered the
effects of the near-field torque.

\subsection{Precession dynamics in the inertial frame}

\begin{figure}
	\centering
	\includegraphics[width=8cm]{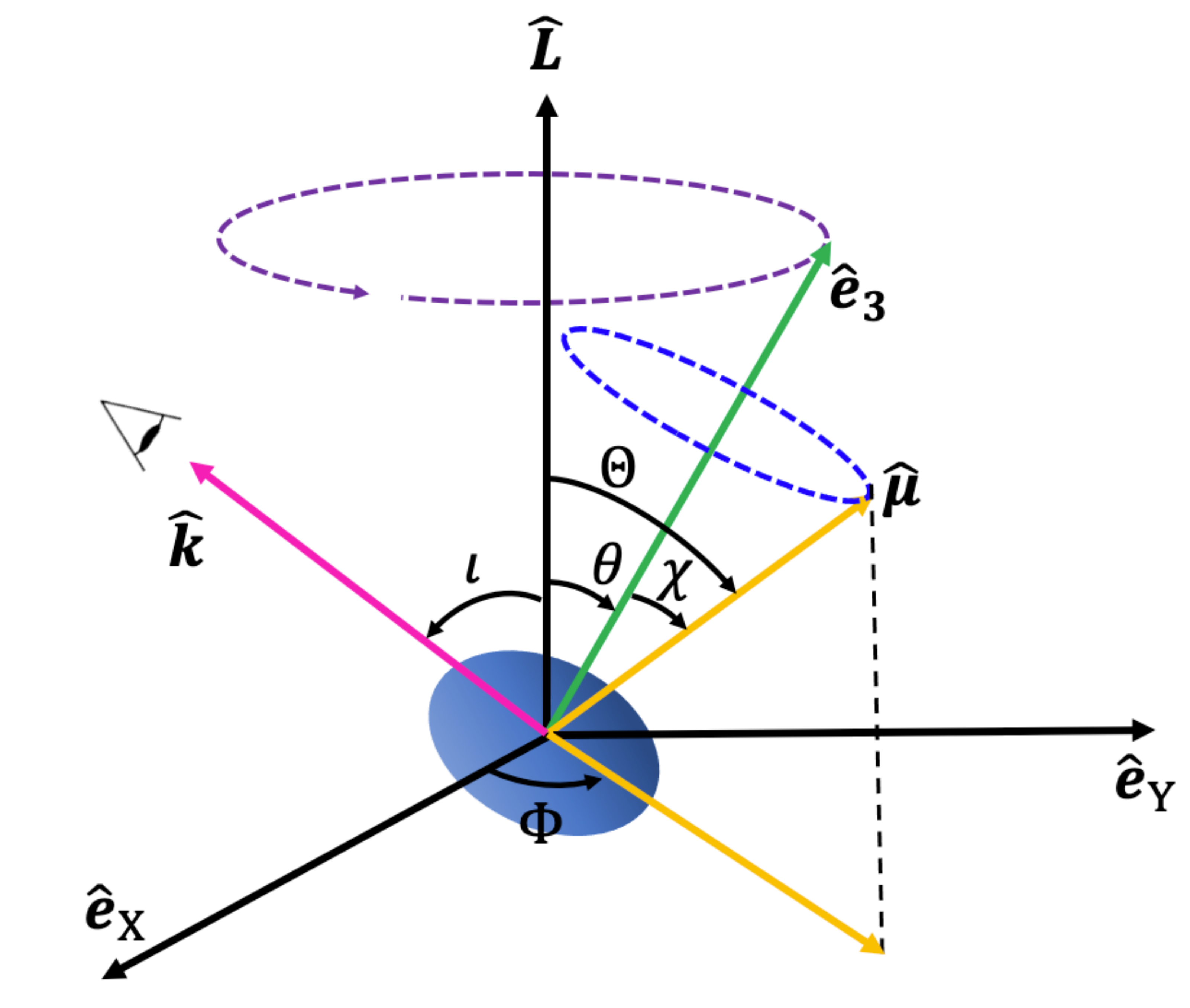}
	\caption{The geometry and the motion of a NS in the inertial frame. The NS
	rotates around $\hat{\boldsymbol{L}}$ with angular frequency
	$\boldsymbol{\omega}$.  The NS itself rotates around
	$\hat{\boldsymbol{e}}_{3}$ with free precession period $T$, which is
	clockwise in the case of $m<1$ and counterclockwise in the case of $m>1$.
	The wobble angle $\theta$ between $\hat{\boldsymbol{e}}_{3}$ and
	$\hat{\boldsymbol{L}}$ nutates with period $T/2$. For the special biaxial
	case, the wobble angle $\theta$ is fixed. An observer in the
	$\hat{\boldsymbol{e}}_{\rm X}$-$\hat{\boldsymbol{e}}_{\rm Z}$ plane views
	the NS in the direction $\hat{\boldsymbol{k}}$ with an inclination angle
	$\iota$. The magnetic dipole ${\boldsymbol{\mu}}$ is attached on the NS and
	is described by the polar angle $\Theta$ and azimuthal angle $\Phi$.}
	\label{fig:inertial_frame}
\end{figure}

The calculations in Secs.~\ref{sec:free_precession} to \ref{sec:forced_precession}
are performed in the body frame of the NS.
Before investigating the emissions from precessing magnetars, we give the
geometry and the motion of the NS in the inertial frame, which is related to the
body frame through a rotation matrix constructed from Euler angles $\phi$,
$\theta$, and $\psi$ \citep{landau1960course}.  We take the basis of the
inertial frame as $\hat{\boldsymbol{e}}_{\rm X}$, $\hat{\boldsymbol{e}}_{\rm
Y}$, and $\hat{\boldsymbol{e}}_{\rm Z}$ with $\hat{\boldsymbol{L}}$ parallel to
$\hat{\boldsymbol{e}}_{\rm Z}$. The Euler angles satisfy 
\begin{align}
	\cos \phi &=\hat{e}_{\mathrm{X}} \cdot \hat{\boldsymbol{N}}, \nonumber \\
	\cos \theta &=\hat{\boldsymbol{e}}_{3} \cdot \hat{e}_{\mathrm{Z}}, 
	\nonumber\\ 
	\cos \psi &=\hat{\boldsymbol{e}}_{1} \cdot \hat{\boldsymbol{N}}\,,
\end{align}
where $\hat{\boldsymbol{N}}=\hat{\boldsymbol{e}}_{Z} \times
\hat{\boldsymbol{e}}_{3}$. The time evolution of the Euler angles are given by 
\begin{align}
	\cos\theta &=\hat{{L}}_{3}\,,\nonumber \\
	\tan \psi  &= \hat{{L}}_{1}/\hat{{L}}_{2}\,,\nonumber \\
	\dot{\phi} &= L/I_{3}-\dot{\psi}/\hat{{L}}_{3}\,.
	\label{eqn:Euler_angles}
\end{align}
Substituting the evolution of the angular momentum for different cases into
Eqs.~(\ref{eqn:Euler_angles}), one obtains the specific expressions for Euler
angles. For the general triaxial case, the precession angle $\psi$ and the
wobble angle $\theta$ evolve with free precession period $T$, while the
evolution of the angle $\phi$ is not periodic. Thus, the motion for a triaxial
NS in the inertial frame is not periodic. We illustrate the geometry and the
motion of the NS in the inertial frame in Fig.~\ref{fig:inertial_frame}.

The components of $\hat{\boldsymbol{\mu}}$ in the inertial frame are 
\begin{align}
	\hat{\mu}_{\rm X}&=  \hat{\mu}_{1}(\cos \psi \cos \phi-\cos \theta \sin \phi \sin \psi) \nonumber\\
	& \quad -\hat{\mu}_{2}(\sin \psi \cos \phi+\cos \theta \sin \phi \cos \psi)+\hat{\mu}_{3} \sin \theta \sin \phi \nonumber\,,\\
	\hat{\mu}_{\rm Y}&=\hat{\mu}_{1}(\cos \psi \sin \phi+\cos \theta \cos \phi \sin \psi)\nonumber \\
	& \quad +\hat{\mu}_{2}(-\sin \psi \sin \phi+\cos \theta \cos \phi \cos \psi)-\hat{\mu}_{3} \sin \theta \cos \phi    \nonumber\,,\\
	\hat{\mu}_{\rm Z}&=\hat{\mu}_{1} \sin \theta  \sin \psi +\hat{\mu}_{2} \sin \theta  \cos \psi +\hat{\mu}_{3} \cos \theta   \,,
	\label{eqn:mu}
\end{align}
where $\hat{\mu}_{1}$, $\hat{\mu}_{2}$, and $\hat{\mu}_{3}$ are the components
of $\hat{\boldsymbol{\mu}}$ in the body frame in Eq.~(\ref{eqn:dipole_body}).
The polar angle $\Theta$ and the azimuthal angle $\Phi$ of the magnetic dipole
in the inertial frame satisfy
\begin{align}
	\Phi &= \arctan \left(\frac{\hat{\mu}_{\rm Y}}{\hat{\mu}_{\rm X}}\right)\,,\nonumber \\
	\cos\Theta &=\hat{\mu}_{\rm Z}\,.
		\label{eqn:Phi_Theta}
\end{align}
We can treat $\Theta$ as the magnetic inclination angle $\alpha$ because the
angle between $\hat{\boldsymbol{L}}$ and $\hat{\boldsymbol{\omega}}$ is 
first order in $\epsilon$. 

The time evolution of $\hat{\boldsymbol{\mu}}$ is vital to determine the
emission properties. The variations of the angle $\alpha$ during free precession
lead to the swing of the emission regions and may modulate the beam shape
parameters, polarization, and flux of the emission. While the precession also
affects the rotational phase and time of arrivals of the emissions, which are
closely related to the time evolution of $\Phi$. In the following sections, we will
first investigate the phase modulations and timing residuals buried in the time
evolution of $\Phi$, and then study the modulations of polarized radio/X-ray
signals due to the variations of $\alpha$ during precession. Note that the calculations 
will be performed in the inertial frame.

\section{Timing residuals}
\label{sec:timing}

In this section, we investigate the timing residuals of precessing magnetars,
which may be used to search for precession from X-ray pulsations.  The main
manifestation of magnetars occurs in the X-ray energy band. Some magnetars are
persistent X-ray sources with a luminosity $L_{\rm X}\sim
10^{34}$--$10^{35}\,\rm erg\,\rm s^{-1}$ while others are transient that are
much dimmer in quiescence, $L_{\rm X}\lesssim 10^{32}\,\rm erg\,\rm
s^{-1}$~\citep{Turolla:2015mwa,Makishima2016,Kaspi:2017fwg}. Most magnetars show
clear X-ray pulsations due to the spin. The periods are clustered in the range
$P=2$--$12\,\rm s$. Most of them show a large spin-down rate in the range of
$\dot{P}= 10^{-13}$--$10^{-10}\,\rm s\,s^{-1}$.  The timing of radio signals are
also obtained for some transient magnetars. 

The rotation phase and spin rate will be modulated since the emission direction
rotates around $\hat{\boldsymbol{e}}_{3}$ during the precession. For simplicity,
we assume that the emission is centered around the magnetic dipole axis
$\hat{\boldsymbol{\mu}}$.  As shown in Fig.~\ref{fig:inertial_frame}, the
observer sees the pulsation once the azimuthal angle of $\hat{\boldsymbol{\mu}}$
becomes
\begin{equation}
	\label{eqn:modu1}
	\Phi=2\pi n,\quad (n=0, 1, 2, \dots)\,,
\end{equation}
which is equivalent to $\mu_{\rm X}>0$ and $\mu_{\rm Y}=0$. Taking
Eq.~(\ref{eqn:mu}), the Euler angle $\phi$ at this epoch can be written 
as~\citep{Jones:2000ud,Akgun:2005nd}
\begin{equation}
	\label{eqn:phi_pulse}
	\phi=2\pi n +\frac{\pi}{2} + \arctan \phi_{1} \,,
\end{equation}
where 
\begin{equation}
	\tan \phi_{1}=\frac{\hat{\mu}_{1} \cos \psi-\hat{\mu}_{2} \sin \psi}{\hat{\mu}_{2} \cos \theta \cos \psi-\hat{\mu}_{3} 
	\sin \theta+\hat{\mu}_{1} \cos \theta \sin \psi}\,.
\end{equation}
To obtain the timing residual, we first study the effective free precession case
including the near-field torque.  The Euler angle $\phi$ integrating from
Eq.~(\ref{eqn:Euler_angles}) is
\begin{equation}
	\label{eqn:phi_general}
	\phi(t)=\phi_{0}+\frac{L}{I_{3}}t + \frac{\sqrt{1+\delta}\omega_{\rm p}}{\cos\theta_{0}}\int_{0}^{t}\frac{\dd t}{1+\delta 
	\operatorname{sn}^{2}\tau}\,,
\end{equation}
where $\phi_{0}$ is the initial phase of $\phi$ and $\tau=\omega_{\rm
p}t+\psi_{0}$. Combining Eq.~(\ref{eqn:phi_pulse}) and 
Eq.~(\ref{eqn:phi_general}), we obtain the time of arrival (TOA) $t_{\rm n}$ of
the $n$-th pulse 
\begin{equation}
	\frac{L}{I_{3}}t_{\rm n}=2\pi n +\frac{\pi}{2} + \arctan \phi_{1}-\phi_{0}- 
	\frac{\sqrt{1+\delta}\omega_{\rm p}}{\cos\theta_{0}}\int_{0}^{t_{n}}\frac{\dd t}{1+\delta \operatorname{sn}^{2}\tau}\,.
\end{equation}
The TOA contains all the information of the precessing timing behaviours. To
further investigate the spin modulations, we give the residuals of the period
and the period derivatives. In practice, one first obtains the period
$P_{0}=2\pi/\omega_{0}$ at some epoch $t_{0}$, and finds the period derivative
$\dot{P}_{0}$ that is attributed to the secular spin down. Then the period 
residuals can be determined by subtracting the two contributions 
\begin{equation}
	\Delta P = P(t)-P_{0}(t_{0})-\dot{P}_{0}(t-t_{0})\,.
\end{equation}

Since the precession timescale is much longer than the rotation timescale, we
can approximate the differences by derivatives, and 
\begin{align}
	\frac{L}{I_{3}} P-2 \pi &= \frac{L}{I_{3}} \Delta P_{\rm fp}\nonumber \\
	&=\left(\frac{\dd \arctan\phi_{1}}{\dd t}-\frac{\sqrt{1+\delta}\omega_{\rm p}/\cos\theta_{0}}{1+\delta 
	\operatorname{sn}^{2}\tau}\right)\frac{\dd t}{\dd n}\nonumber \\
	&= \left(\frac{\dd \arctan\phi_{1}}{\dd t}-\frac{\sqrt{1+\delta}\omega_{\rm p}/\cos\theta_{0}}{1+\delta 
	\operatorname{sn}^{2}\tau}\right)P\,,
\end{align} 
where the period $P=t_{\rm n}-t_{\rm n-1}$, and $\Delta P_{\rm fp}$ is the
period residual owing to the free precession. By approximating $P\simeq P_{0}$
on the right-hand side, we get
\begin{equation}
	\Delta P_{\rm fp}=\left(\frac{\dd \arctan\phi_{1}}{\dd \tau }-\frac{\sqrt{1+\delta}/\cos\theta_{0}}{1+
	\delta \operatorname{sn}^{2}\tau}\right)\frac{\epsilon\cos\theta_{0}P_{0}}{\sqrt{1+\delta}}\,.
\end{equation}
For an effectively biaxial case or a biaxial case with $\epsilon_{m}\ll
\epsilon$, we can set $\hat{\mu}_{2}=0$ and 
\begin{align}
	\Delta P_{\rm fp} =& -\epsilon \sin\theta_{0}P_{0}\nonumber\\
	 & \times\biggl[\frac{\mu_{1}^{2} \sin \theta_{0} \sin ^{2} \omega_{\rm p} t+\mu_{3}^{2} \sin \theta_{0}-\mu_{1} \mu_{3} 
	 \cos \theta_{0} \cos \omega_{\rm p} t}{\left(\mu_{1} \cos \theta_{0} \cos \omega_{\rm p} t-\mu_{3} 
	 \sin \theta_{0}\right)^{2}+\mu_{1}^{2} \sin ^{2} \omega_{\rm p} t}\biggr]\,,
\end{align}
where the initial phase induced by the near-field torque $\psi_{0}=0$. Because
$\Delta P_{\rm fp}$ purely originates from the geometry of the free precession,
we name $\Delta P$ as the geometric term of  residual in period.

The far-field torque contributes to the period residual via the spin down 
\begin{equation}
	\frac{\Delta P_{\rm sd}}{P}\simeq -\ell(t)\,,
\end{equation}
where $\ell(t)$ has been given in Eq.~(\ref{eqn:fractional}) with the
integration of $\cos\alpha$ in Eq.~(\ref{eqn:int_alpha}).  For period residuals,
we only care about the oscillation terms. After subtracting the secular terms,
the period residual is 
\begin{align}
	\Delta P_{\rm sd}&=-\frac{3k_{1}P_{0}}{2\tau_{\rm rad}}\left(\int_{0}^{t}\cos^{2}\alpha \dd t-\left\langle 
		\int_{0}^{t}\cos^{2}\alpha \dd t \right\rangle t \right) \nonumber\\
	&\approx \frac{3k_{1}P_{0}}{2\tau_{\rm rad}\omega_{\rm p}}\biggl\{a_{1}\operatorname{cn}\tau+a_{2}\operatorname{sn}\tau 
	+ a_{3}\operatorname{dn}\tau \nonumber\\
	& + a_{4}\left[\frac{E(m)}{K(m)}\tau-E\left({\rm a m} \,\tau\right)\right] +B_{\rm c}\biggr\} \,,
\end{align}
where $\left\langle \ell \right\rangle$ means an average over the precession
period and 
\begin{align}
	&a_{1}=\sin 2 \theta_{0}(1+\delta)^{\frac{1}{2}} \hat{\mu}_{2} \hat{\mu}_{3} \,,\nonumber\\
	&a_{2}=-\sin 2\theta_{0} \hat{\mu}_{1} \hat{\mu}_{3} \operatorname{sn} (\tau)\,,\nonumber \\
	&a_{3}= \frac{2 \cos ^{2} \theta_{0} \hat{\mu}_{1} \hat{\mu}_{2}(1+\delta)^{\frac{1}{2}}}{\delta} \,, \nonumber\\
	&a_{4}= \frac{\cos ^{2} \theta_{0}}{\delta}\left[\hat{\mu}_{1}^{2}-(1+\delta) \hat{u}_{2}^{2}+ \hat{\mu}_{3}^{2}\delta\right] \,.
\end{align}
The constant $B_{\rm c}$ is an integration constant, which can be easily obtained
from $\Delta P_{\rm sd}(t=0)=0$.  For the special biaxial case, we set
$\hat{\mu}_{2}=0$ and 
\begin{align}
	\label{eqn:spindown_term_biaxial}
	\Delta P_{\rm sd}=&\ -\frac{3k_{1}P_{0}}{2\tau_{\rm rad}\omega_{\rm p}}\nonumber\\
	& \times \left(\frac{1}{2}\sin2\chi\sin2\theta_{0}\sin(\omega_{\rm p}t) +\frac{1}{4}\sin^{2}\theta_{0}\sin^{2}\chi\sin(2\omega_{\rm p}t)\right)\,,
\end{align}
which is consistent with~\citet{Jones:2000ud} and \citet{Link:2001zr}. We name
the period residual resulting from the far-field torque as the spin-down term.

\def\arraystretch{1.3}
\begin{table*}
	\caption{The intrinsic and effective parameters for the timing residuals shown in Fig.~\ref{fig:timing1}-\ref{fig:timing5}.}
	\centering 
	\begin{tabular}{lccccccccccccc}
		\toprule
		 Case&\multicolumn{7}{c}{Intrinsic parameters}&\multicolumn{6}{c}{Effective parameters} \\
		 \cmidrule(lr){2-8}\cmidrule(lr){9-14}
		 &
		$P_{0}\,(\rm s)$&
		$B\,(\rm G)$ & 
		$\epsilon$ & 
		$\delta$ & 
		$\theta_{0}\, (\degr)$ & 
		$\chi\, (\degr)$ & 
		$\eta \, (\degr) $&
		$\epsilon_{\rm eff}$ &
		$\delta_{\rm eff}$ & 
		$\theta_{\rm eff, 0}\, (\degr)$ & 
		$\chi_{\rm eff}\, (\degr) $ & 
		$\eta_{\rm eff} \, (\degr) $ & 
		$T_{\rm eff}\,(\rm yr)$ \\ 
		\hline
		\uppercase\expandafter{\romannumeral1} & 5 & $ 10^{14}$ & $10^{-7}$ & 0 & 15&  45 & 0 & $1.00\times 10^{-7}$& $7.61\times10^{-3}$& 15.4&45.4 & 0 & 1.65 \\
	
		\uppercase\expandafter{\romannumeral2} & 5 & $ 10^{14}$ & $10^{-7}$ & 1 & 15&  45 & 45 & $9.96\times 10^{-8}$& 1.01& 15.3 & 45.6 & 44.8 & 2.38 \\
		
		\uppercase\expandafter{\romannumeral3} & 5 & $ 10^{14}$ & $10^{-7}$ & 0 & 15&  85 & 0 & $1.01\times 10^{-7}$& 0.0149& 15.1& 85.1 & 0 & 1.63 \\
		
		\uppercase\expandafter{\romannumeral4} & 5 & $ 10^{14}$ & $10^{-7}$ & 1 & 15&  85 & 45 & $1.01\times 10^{-7}$& 0.986& 15.1&85.1 & 44.2 & 2.34 \\

		\uppercase\expandafter{\romannumeral5} &{5}& $ {10^{14}}$ & ${10^{-4}}$ & {0} & {15}&  {45} & {0 }& ${1.00\times 10^{-4}}$& ${7.50\times10^{-6}}$& {15.0}&{45.0}& {0 }& {0.00164}\\

		\uppercase\expandafter{\romannumeral6} & {5} & $ {10^{14}}$ & {$10^{-4}$}& {1} & {15}&  {45} & {45} & ${1.00\times 10^{-4}}$& {1.00}& {15.0} & {45.0} & {45.0} & {0.00236} \\

		\uppercase\expandafter{\romannumeral7} & 5 & $5\times 10^{14}$ & $10^{-7}$ & 0 & 15&  45 & 0 & $1.07\times 10^{-7}$& 0.261& 25.3&55.3 & 0 & 1.87 \\

		\uppercase\expandafter{\romannumeral8} & 5 & $5\times 10^{14}$ & $10^{-7}$ & 1 & 15&  45 & 45 & $9.99\times 10^{-8}$& 1.15& 24.9&60.2 & 36.1 & 2.75 \\

		\uppercase\expandafter{\romannumeral9} & {5} & $ {10^{14}}$ & ${10^{-5}}$ & {0} & {15}&  {45} & {0} & ${1.00\times 10^{-5}}$& ${7.50\times10^{-5}}$& {15.0}&{45.0} & {0}& {0.0164}\\

		\uppercase\expandafter{\romannumeral10} & {5 }& $ {10^{14}}$ & ${10^{-5}}$ &{8} & {15}&  {45} & {0} & ${1.00\times 10^{-5}}$& {8.01}& {15.0} & {45.0} & {45.0} & {0.0603}\\

		\bottomrule
	\end{tabular}
	\label{tab:second}
\end{table*}


\begin{figure}
	\centering
	\includegraphics[width=8cm]{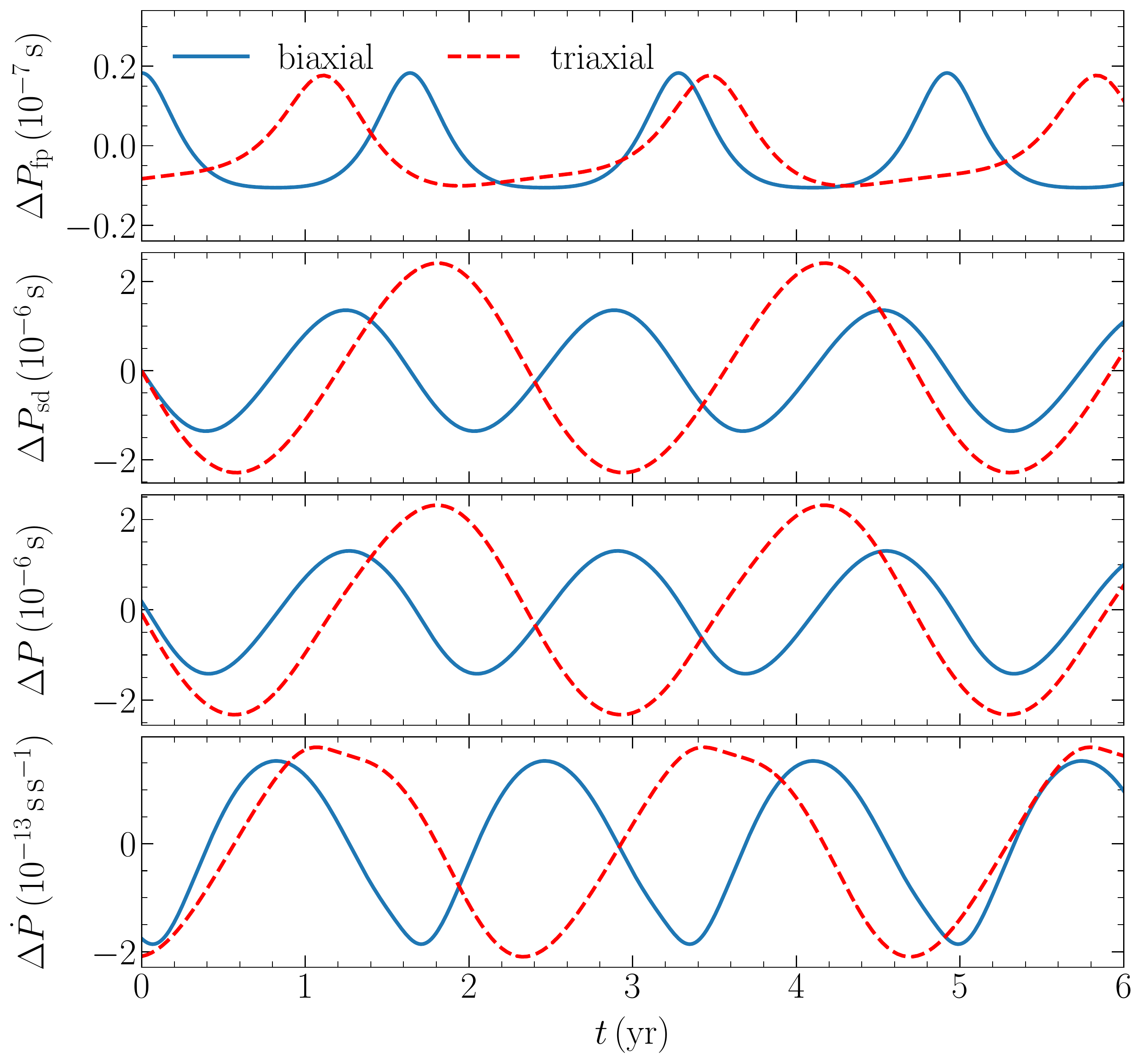}
	\caption{The residuals of the period and the period derivative for biaxial
	and triaxial NSs with $k_{1}=1$ and $k_{2}=2$. The parameters for the biaxial and triaxial 
	cases are shown in Case \uppercase\expandafter{\romannumeral1} and Case 
	\uppercase\expandafter{\romannumeral2} of Table~{\ref{tab:second}} respectively.
	The initial period derivative is
	$\dot{P}_{0}=8.22\times 10^{-13}\,\rm s\,s^{-1}$ for the biaxial case and
	$\dot{P}_{0}=8.82\times 10^{-13}\,\rm s\,s^{-1}$ for the triaxial case.}
	\label{fig:timing1}
\end{figure}

\begin{figure}
	\centering
	\includegraphics[width=8cm]{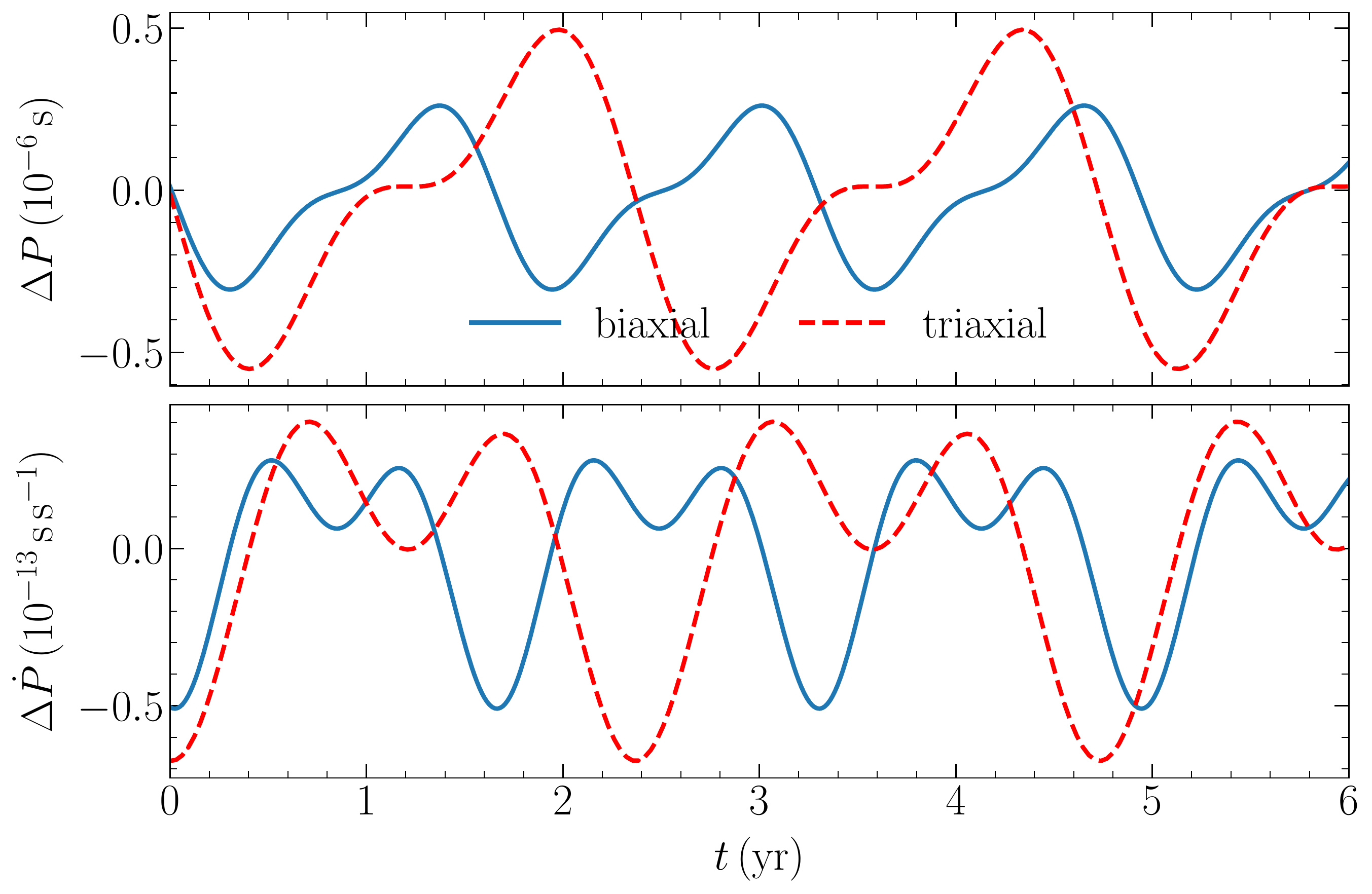}
	\caption{Same as Fig.~\ref{fig:timing1}, except for $\chi=85^{\circ}$. 
	The parameters for the biaxial and triaxial 
	cases are shown in Case \uppercase\expandafter{\romannumeral3} and Case 
	\uppercase\expandafter{\romannumeral4} of Table~{\ref{tab:second}} respectively.
	The initial period derivative is
	$\dot{P}_{0}=1.24\times 10^{-12}\,\rm s\,s^{-1}$ for the biaxial case and
	$\dot{P}_{0}=1.27\times 10^{-12}\,\rm s\,s^{-1}$ for the triaxial case.}
	\label{fig:timing2}
\end{figure}

\begin{figure}
	\centering
	\includegraphics[width=8cm]{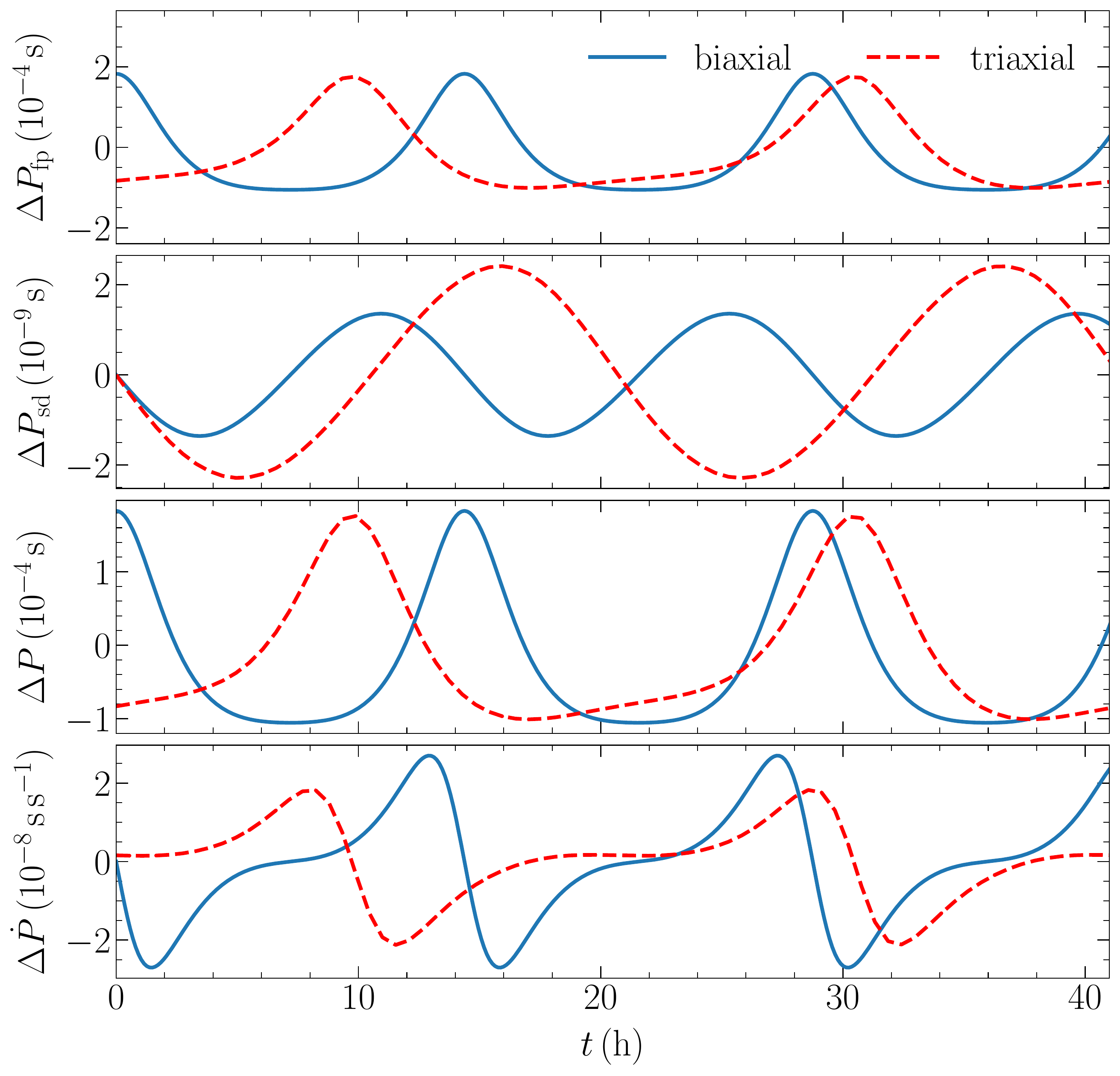}
	\caption{Same as Fig.~\ref{fig:timing1}, but with $\epsilon=10^{-4}$.
	The parameters for the biaxial and triaxial 
	cases are shown in Case \uppercase\expandafter{\romannumeral5} and Case 
	\uppercase\expandafter{\romannumeral6} of Table~{\ref{tab:second}} respectively.
	The initial period derivative is
	$\dot{P}_{0}=8.22\times 10^{-13}\,\rm s\,s^{-1}$ for the biaxial case and
	$\dot{P}_{0}=8.82\times 10^{-13}\,\rm s\,s^{-1}$ for the triaxial case.}
	\label{fig:timing3}
\end{figure}

\begin{figure}
	\centering
	\includegraphics[width=8cm]{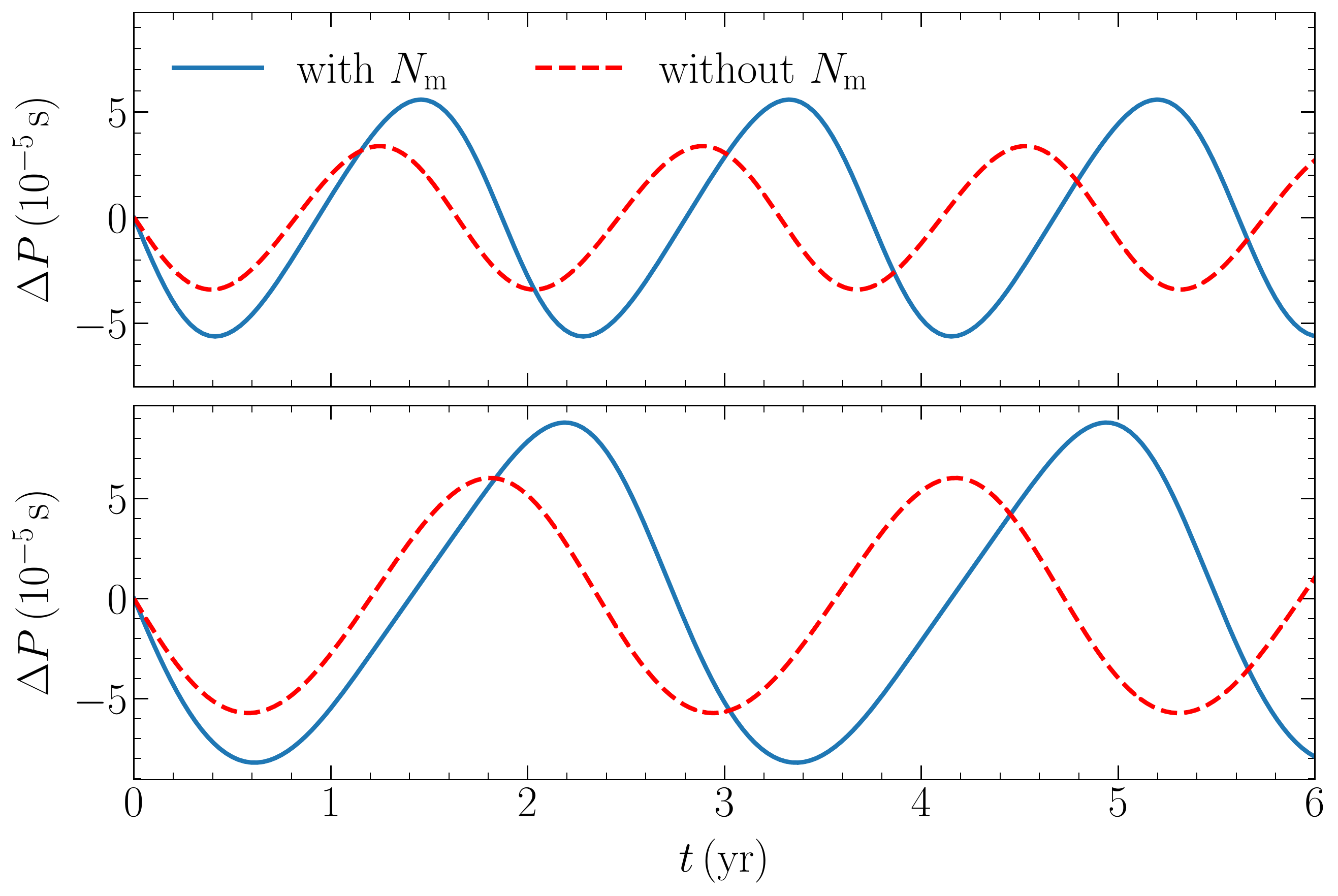}
	\caption{The period residual for a biaxial case (upper) and a triaxial case
	(lower) for $B=5\times 10^{14}\,\rm G$.  
	The parameters for the biaxial and triaxial cases are shown in Case 
	\uppercase\expandafter{\romannumeral7} and Case \uppercase\expandafter{\romannumeral8} 
	of Table~{\ref{tab:second}} respectively. The period derivative
	$\dot{P}_{0}=2.06\times 10^{-11}\rm s\,s^{-1}$ for the biaxial case and
	$\dot{P}_{0}=2.20\times 10^{-11}\rm s\,s^{-1}$ for the triaxial case.}
	\label{fig:timing4}
\end{figure}

\begin{figure}
	\centering
	\includegraphics[width=8cm]{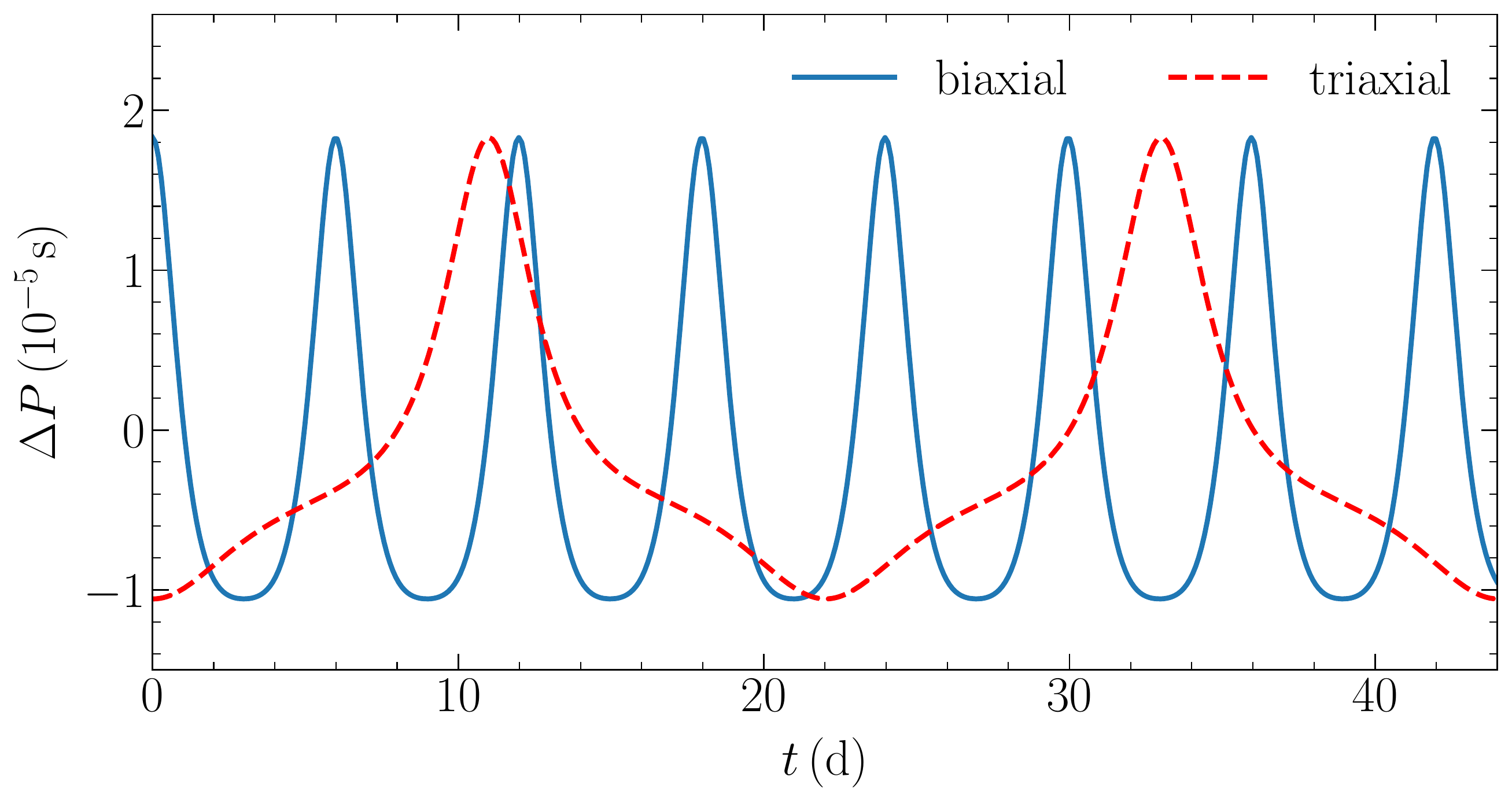}
	\caption{The residuals of the period for biaxial
	and triaxial NSs. The parameters are shown in Case \uppercase\expandafter{\romannumeral9} and Case 
	\uppercase\expandafter{\romannumeral10} of Table~{\ref{tab:second}} respectively.
	The initial period derivative is
	$\dot{P}_{0}=8.22\times 10^{-13}\,\rm s\,s^{-1}$ for the biaxial case and
	$\dot{P}_{0}=8.82\times 10^{-13}\,\rm s\,s^{-1}$ for the triaxial case.}
	\label{fig:timing5}
\end{figure}

The total period residual $\Delta P$ can be expressed as 
\begin{equation}
	\Delta P=\Delta P_{\rm fp} + \Delta P_{\rm sd}\,.
\end{equation}
Here the geometric term can be obtained from the effectively free precession
problem. While the spin-down term is determined by the far-field torque. The
relative amplitude of the two terms depends on the geometry of the star. One
notices that 
\begin{align}
	\frac{\Delta P_{\rm fp}}{P_{0}} &\sim  {\rm coefficient} \times \frac{P_{0}}{\tau_{\rm f}}\,,\nonumber \\
	\frac{\Delta  
	P_{\rm sd}}{P_{0}} &\sim   {\rm coefficient} \times\frac{\tau_{\rm f}}{\tau_{\rm rad}}\,,
\end{align}
where the coefficients are some geometric factors depending on the geometry of
the deformed magnetars. When the free precession timescale $\tau_{\rm f}$ is
sufficiently long, corresponding to small ellipticities, the spin-down term
dominates over the geometric term. This is the case for the possible precession of
PSR~B1828$-$11~\citep{Link:2001zr,Akgun:2005nd}. If the free precession 
timescale $\tau_{\rm f}$ is closer to the spin period other than the spin-down
timescale $\tau_{\rm rad}$, corresponding to large ellipticities, the geometric
term will dominate. This is the case for the possible precession of 4U
0142$+$61~\citep{Makishima:2014dua}.

When $m>1$, the ellipticity is negative. Thus the precession direction is
opposite to the case of $m<1$. If we change $\epsilon$ into $-\epsilon$ and keep
the other parameters fixed, the geometric term changes sign while the spin-down term
stays the same. The amplitude of the spin-down term is proportional to $k_{1}$.
So the amplitude of the period residual due to vacuum torque is just $2/3$ times
that of the plasma-filled torque. In following examples, we only study the
timing residuals of the case with $k_{1}=1$.

To study the effects of the NS geometries and the near-field torque separately,
we first neglect the contributions of the near-field torque by taking
$\epsilon=10^{-7}$ and $B=10^{14}\,\rm G$, where $\epsilon_{\rm m}=0.015\epsilon
\ll \epsilon$.  In Fig.~\ref{fig:timing1}, we give an example of $\Delta P_{\rm
sd}\gg \Delta P_{\rm fp}$. The spin-down term dominates over the geometric term
by a factor of $\sim 10$. The morphologies for the biaxial case and the triaxial
case are basically the same. The main differences are the amplitude and the
period of the modulations. 

Another important feature is that $\Delta P$ does not deviate much from a single
harmonic.  We take the biaxial case as an example to understand this point. The
triaxial case can be understood in the same way qualitatively.  For the biaxial
case, the spin-down term $\Delta P_{\rm sd}$ has components both at frequencies
$\omega_{\rm p}$ and $2\omega_{\rm p}$.  The amplitude of  $\Delta P_{\rm sd}$
at $\omega_{\rm p}$ is larger than the harmonics at $2\omega_{\rm p}$ only if
$\cot \theta > \tan \chi/8$.  For the biaxial case in Fig.~\ref{fig:timing1},
$\cot\theta=\cot \theta_{0}=3.73$ while $\tan \chi/8=0.125$. The residuals at
$\omega_{\rm p}$ is about 30 times larger. Therefore, the residuals mainly come
from the term at the frequency $\omega_{\rm p}$ of the spin-down contribution. 

For the biaxial case, the first harmonic at $2\omega_{\rm p}$ of the spin-down
term only plays an important role when $\cot \theta < \tan \chi/8$.  So we
present an example with $\chi=85^{\circ}$ in Fig.~\ref{fig:timing2} and keep
other parameters fixed. One can notice that the residuals are quite different
from that in Fig.~\ref{fig:timing1} due to that the contribution at
$2\omega_{\rm p}$ is comparable with that at $\omega_{\rm p}$. While the
geometric term is still about 0.1 of the spin-down term. The precession of
PSR~B1828$-$11 belongs to this kind.

To make the geometric term dominate over the spin-down term, the ellipticity
should be sufficiently large. In Fig.~\ref{fig:timing3}, we show an example with
$\epsilon=10^{-4}$ and keep the other parameters the same as Fig.~\ref{fig:timing1}.
The geometric term is much larger than the spin-down term by a factor $\sim
10^{5}$.  The period residual is quite substantial and the effects of the electromagnetic 
torques are negligible. 

Actually, such kind of modulations have been possibly observed by combined
timing analysis of hard and soft X-rays for three magnetars, 
4U~0142+61~\citep{Makishima:2014dua},
1E~1547.0$-$5408~\citep{Makishima:2021vvv}, and
SGR~1900+14~\citep{Makishima:2021hho}. 
\citet{Makishima:2014dua,Makishima:2021vvv,Makishima:2021hho} gave the phase
modulations of hard X-rays, which are physically equivalent to the timing
residuals.  In their model, the internal strong toroidal magnetic field creates
a large prolate deformation along the magnetic dipole.  The soft X-ray emission
is centered around the magnetic dipole while the hard X-ray emission is somewhat
misaligned with the magnetic dipole.  This model is different from ours but can
be simply obtained by redefining $\hat{\boldsymbol{\mu}}$ as the emission
direction of the hard X-rays in a direction other than the magnetic dipole and treating the star
as a biaxial one. Thus, the period residual for 4U 0142$+$61 should be in the
order of $\epsilon P\sim 0.001\,\rm s$ according to
Eq.~(\ref{eqn:spindown_term_biaxial}).  For the magnetar 4U 0142$+$61,
\citet{Makishima:2014dua} found that the rotation period at $8.69\,\rm s$
suffers slow phase modulations of $0.7\,\rm s$, with a period of $\sim 15\,\rm
h$ in the hard X-ray band ($15$--$40\,\rm keV$), indicating an internal magnetic
deformation $\epsilon\sim -1.6\times 10^{-4}$ if the modulations are interpreted
as free precession. 

For the examples in Figs.~\ref{fig:timing1}--\ref{fig:timing3}, the near-field
torque can be neglected. Thus, in Fig.~\ref{fig:timing4}, we show biaxial and
triaxial examples with a large near-field torque. The amplitude of the residuals
becomes larger compared to the cases without the near-field torque. The period
of the modulations turns into $T_{\rm eff}$. 

{In all above examples, the parameter $m$ is on the order of $0.1$. Although the period and amplitude 
of the residuals for the triaxial cases are different from the biaxial ones, the morphologies are 
basically the same. It is easier to tell whether a NS is triaxial or not from the timing residuals
if the parameter $m$ is much larger. In Fig.~\ref{fig:timing5}, we show a triaxial example 
with $\delta=8$, $\theta_{0}=15{\degr}$, and $m=0.574$.  Since the wobble angle nutates in a wider 
range and the Jacobi elliptic functions deviate from the harmonic functions substantially, 
the morphology and period of the timing residuals for the triaxial case are quite different from the biaxial one.}

\section{Modulations on polarizations}
\label{sec:polarization}

The Stokes parameters for the polarizations directly reflect the magnetic field
rotating around the NS and the emission geometry.  The precession leads to the
swing of the emission and changes the polarization. In this section, we model
the polarization of precessing magnetars and study the prospects of detecting the
free precession with polarized X-ray and radio emissions.

\subsection{Emission model of X-rays}

We use the model developed by \citet{Ho:2002jn}, \citet{Lai:2003nd}, and
\citet{vanAdelsberg:2006uu} to calculate the soft thermal X-ray emission from 
the surface of a NS. We assume that the emission comes from a hot region, which
is centered around the magnetic dipole axis, much smaller than the surface area
of the star, and composed of a fully ionized hydrogen atmosphere with an
effective temperature $T_{\rm eff}\simeq 5\times 10^{6}\,\rm K$. The magnetic
field is also assumed to be a dipole field, which is approximately constant and
normal to the stellar surface across the emission region. 

In the highly magnetized plasma that characterizes the magnetosphere of NSs,
X-ray photons propagate in the extraordinary mode (X mode) and the ordinary mode
(O mode). The X mode is mostly polarized perpendicular to the
$\boldsymbol{k}_0$-$\boldsymbol{B}$ plane while the ordinary mode (O mode) is
mostly polarized within the $\boldsymbol{k}_0$-$\boldsymbol{B}$ plane, where
$\boldsymbol{k}_{0}$ is the direction of photon propagation direction at the
emission point and $\boldsymbol{B}$ is the external magnetic field. The
opacities for each mode are associated with the energy and the propagation
direction of the X-ray photons, as well as the strength and the direction of the
magnetic field in the magnetized plasma. The typical X mode opacity $\kappa_{\rm
X}$ is much smaller than the O mode opacity $\kappa_{\rm
O}$~\citep{Meszaros1992}, satisfying  $\kappa_{\rm X} \sim\left(E / E_{B
e}\right)^{2} \kappa_{\rm O}$, where $E_{\rm Be}=\hbar e B/m_{\rm e} c$ is the
electron cyclotron energy in the magnetic field. The decoupling density of the X
mode photon $\rho_{\rm X}$ is much larger than that of the O mode photon
$\rho_{\rm O}$. As a result, the X mode photons can escape from deeper and
hotter layers than the O mode photons.  The emergent radiation is linearly
polarized to a high
degree~\citep{Gedin1974,Meszaros:1988ns,Pavlov:1999cw,Ho:2002jn,Lai:2003nd}. 

In strong magnetic fields, it has long been predicted that the vacuum becomes
birefringent, and the dielectric tensor describing the atmospheric plasma of
magnetars must be corrected for quantum electrodynamics (QED) vacuum
effects~\citep{Heisenberg:1936nmg,Tsai:1975iz}.  At the vacuum resonance, the
contributions of the plasma and the vacuum to the dielectric tensor cancel each
other \citep{Lai:2003nd}.  When a photon with energy $E$ traverses through the
density gradient of the plasma, it will encounter the vacuum resonance at the
density 
\begin{equation}
	\rho_{\rm V}=0.96 Y_{e}^{-1} E_{1}^{2} B_{14}^{2} f_{\rm B}^{-2} \mathrm{~g} \mathrm{~cm}^{-3}\,,
\end{equation}
where $Y_{e}=Z/A$ with Z and A  the charge number and mass number of the ion
respectively, $E_{1}=E/(1\,\rm keV)$, and $f_{\rm B}$ is a slowly varying
function of $B$ that is on the order of unity. If the density variations of the
plasma are sufficiently gentle as the photon propagates through the
inhomogeneous plasma, an X mode (O mode) photon will be converted into an O mode
(X mode) photon when it traverses the vacuum resonance. For the mode conversion
to be effective, the adiabatic condition $E\geq E_{\rm ad}$ must be
satisfied~\citep{Ho:2002jn,Lai:2003nd}, with 
\begin{equation}
 	E_{\mathrm{ad}}=2.52\left[f_{\rm B} \tan \theta_{\rm k B}\left|1-\left(E_{\rm B i} / 
	E\right)^{2}\right|\right]^{2 / 3}\left(\frac{1 \mathrm{~cm}}{H_{\rm \rho}}\right)^{1 / 3}\,.
\end{equation}
Here $\theta_{\rm kB}$ is the angle between the magnetic field and the photon
propagation direction, $E_{\rm Bi}$ is the ion cyclotron energy, and $H_{\rm
\rho}$ is the density scale-height along the ray. For a photon with energy
$E\sim E_{\rm ad}$, it undergoes partial mode conversion.  In general, the mode
conversion probability of a photon is~\citep{Lai:2003nd}
\begin{equation}
 	P_{\rm c}=1-\exp \left[-(\pi / 2)\left(E / E_{\mathrm{ad}}\right)^{3}\right]\,.
\end{equation}

To obtain the emergent intensities, one needs to solve the radiative transfer
equations (RTEs) of the two modes subject to the constraints of hydrostatic
and radiative equilibria~\citep{Ho:2001ik,Ho:2002jn}.
\citet{vanAdelsberg:2006uu} provided the fitting formulae of the temperature 
profile for different atmospheric models with different magnetic fields $B$ and
effective temperatures $T_{\rm eff}$. Once the temperature profile is known, the
emergent intensities can be obtained by a single integration of the RTEs. We use the
fitted temperature profile and integrate the RTEs including the vacuum effect
following~\citet{vanAdelsberg:2006uu}. The spectral intensities for the X mode
photons $I_{\rm X}(\theta_{\rm em})$ and the O mode photons $I_{\rm
O}(\theta_{\rm em})$ at different emission angles $\theta_{\rm em}$ are
obtained. The ``intrinsic'' linear polarization fraction at the emission point
is defined as
\begin{equation}
	\Pi_{\rm em}=\frac{I_{\rm O}(\theta_{\rm em})-I_{\rm X}(\theta_{\rm em})}
	{I_{\rm O}(\theta_{\rm em})+I_{\rm X}(\theta_{\rm em})}\,,
\end{equation}
where $\theta_{\rm em}$ is the angle between the photon propagation direction
and the surface normal.

\begin{figure}
	\centering
	\includegraphics[width=8cm]{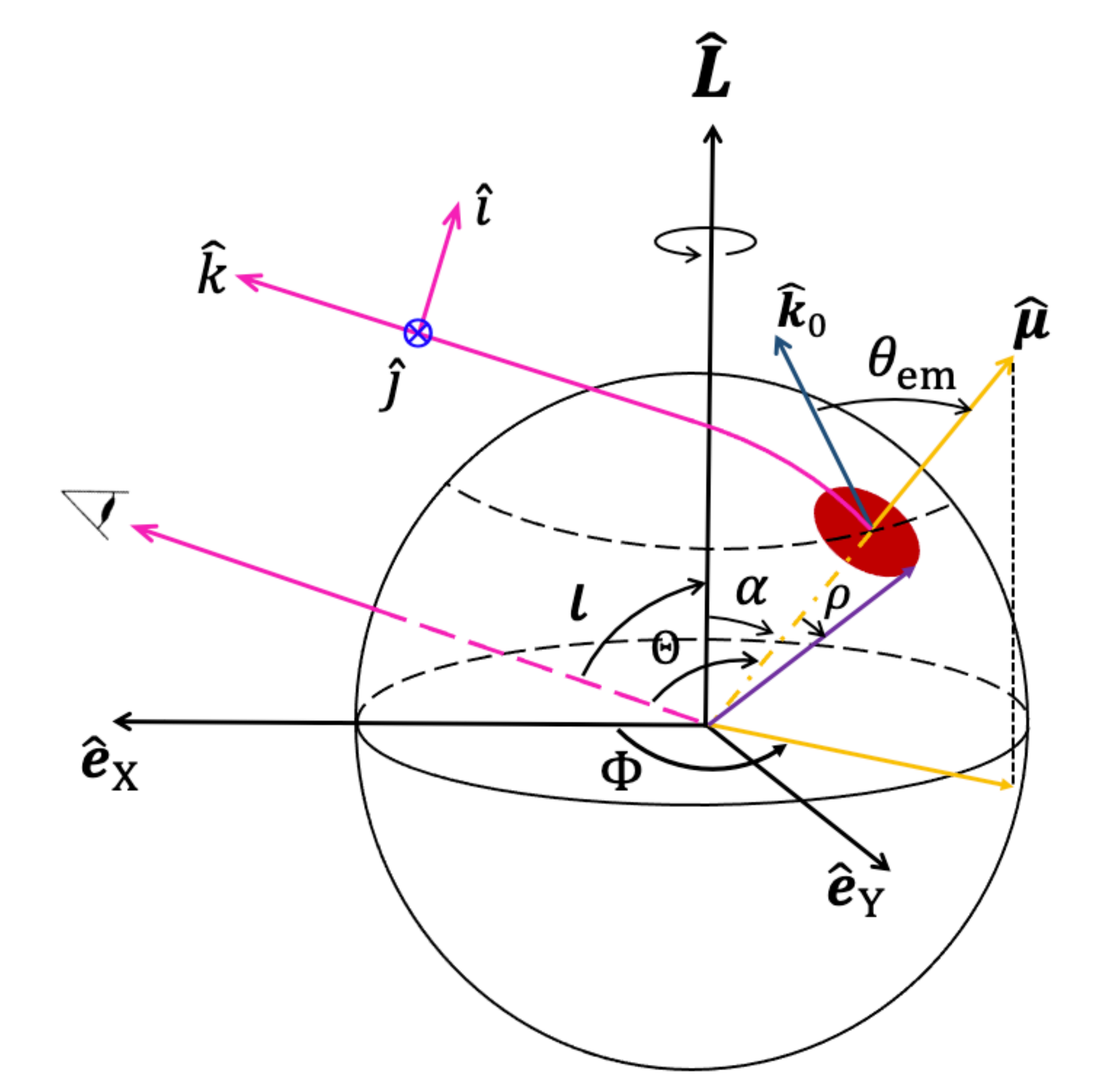}
	\caption{The X-ray emission geometry. The observer lies in the
	$\hat{\boldsymbol{e}}_{\rm X}$-$\hat{\boldsymbol{e}}_{\rm Z}$ plane at an
	inclination angle $\iota$. A hotspot is located at one of the magnetic
	poles. An X-ray photon emitted at an angle $\theta_{\rm em}$ respect to the
	surface normal will be received at colatitude $\Theta$ due to the light
	bending effect. A coordinate system $IJK$ with the basis $\{\hat{\boldsymbol{i}},
	\hat{\boldsymbol{j}}, \hat{\boldsymbol{k}}\}$ is introduced, where
	$\hat{\boldsymbol{k}}$ is along the line of sight, $\hat{\boldsymbol{i}}$
	lies in the plane spanned by the line of sight and the angular momentum
	$\boldsymbol{L}$, and $\hat{\boldsymbol{j}}$ is determined by
	$\hat{\boldsymbol{L}}\times \hat{\boldsymbol{k}}=- \hat{\boldsymbol{j}}
	\sin\iota  $.}
	\label{fig:X-ray}
\end{figure}

To determine the polarization state of the signals, one must consider the
propagation of polarized radiation in the magnetosphere of magnetars, whose
dielectric properties are dominated by the vacuum polarization in the X-ray
band~\citep{Heyl:2003kt}. When an X-ray photon propagates in the magnetosphere,
its polarization state evolves adiabatically along the varying magnetic field up
to the polarization limiting radius $r_{\rm pl}$, which is far from the surface
of the NS. Thus, the observed Stokes parameters are determined by the ``frozen''
polarization state at $r_{\rm pl}$. Adiabatic evolution of the photon modes in
the magnetosphere leads to a significant polarization fraction even when the
emission comes from extended regions on the
surface~\citep{Heyl:2003kt,Fernandez:2011aa, Taverna:2015vpa}. In contrast, if
the polarization state is determined by the emission at the surface ,
additions of the Stokes parameters from distinct regions tend to cancel each
other and lead to low polarization fraction.

The magnetic field direction that determines the polarization can be
characterized by the polar angle $\Theta$ and the azimuthal angle $\Psi$. As
shown in Fig.~\ref{fig:X-ray}, the polar angle between the magnetic dipole field
and the line of sight satisfies 
\begin{equation}
	\label{eqn:Theta_Phi}
	\cos\Theta = \cos\iota \cos\alpha  +  \sin\iota \sin\alpha \cos\Phi\,.
\end{equation}
The azimuthal angle $\Psi$ is the position angle (PA) of the polarized emission.
To obtain the PA, we project the dipole field onto the
$\hat{\boldsymbol{i}}$-$\hat{\boldsymbol{j}}$ plane. Introducing the
polarization basis 
\begin{align}
	\hat{\boldsymbol{e}}_{1}^{\rm p} &=\frac{(\hat{\boldsymbol{k}}\times \hat{\boldsymbol{\mu}})\times 
	\hat{\boldsymbol{k}}}{\sin \Theta}
	\,,\nonumber \\
	\hat{\boldsymbol{e}}_{2}^{\rm p} &=\frac{\hat{\boldsymbol{k}}\times 
	\hat{\boldsymbol{\mu}}}{\sin \Theta}\,,
\end{align}
the PA measured from the projection of the spin axis onto the plane of the sky
in the counterclockwise direction is given by
\begin{align}
	\cos\Psi & =\hat{\boldsymbol{e}}_{1}^{\rm p}\cdot \hat{\boldsymbol{i}}=\frac{\sin \iota \cos \alpha-
	\cos \iota \sin \alpha \cos \Phi}{\sin \Theta}\,,\\
	\sin\Psi & =\hat{\boldsymbol{e}}_{1}^{\rm p}\cdot \hat{\boldsymbol{j}}=-\frac{\sin \alpha \sin \Phi}{\sin \Theta}\,.
\end{align}
Then, we obtain the expressions of PA in the RVM as~\citep{rvm1969,Lorimer:2005misc}
\begin{equation}
	\tan \Psi=\frac{\sin \alpha \sin \Phi}{\cos \iota \sin \alpha \cos \Phi-\sin \iota \cos \alpha}\,.
\end{equation}
The rotation phase at the polarization limiting radius is $\Phi(r_{\rm pl})=
\Phi(R)+r_{\rm pl}/R_{\rm LC}$, where $\Phi(R)$ is the
rotation phase when the photon is emitted at the surface. Magnetars rotate 
slowly, with $r_{\rm pl}/R_{\rm LC}\ll 1$ and $\Phi(r_{\rm pl})\simeq \Phi(R)$. 

In principle, one needs to evolve the polarization state along the magnetic
field to determine $\Psi$ for different points on the extended
hotspot~\citep{Heyl:2003kt,Taverna:2015vpa}. However, we consider a hot region
much smaller than the surface area of magnetars and the magnetic field is
constant across the emission region. Under this condition, the observed PA can
be approximated as $\Psi(r_{\rm pl})\simeq
\pi+\Psi(R)$~\citep{Lai:2003nd,vanAdelsberg:2006uu}. Therefore, the polarization
state only changes with a constant phase shift compared to the intrinsic one.
The Stokes parameters $Q$ and $U$ that are normalized to the total intensity $I$
are 
\begin{align}
	Q/I=&\Pi_{\rm em}\cos 2\Psi(r_{\rm pl})\,,\\
	U/I=&\Pi_{\rm em}\sin 2\Psi(r_{\rm pl})\,.
\end{align}

To obtain the spectral ``flux'' of the Stokes parameters, the propagation
effects in the curved spacetime such as the light bending and gravitational
redshift need to be considered. In the $IJK$ frame shown in
Fig.~\ref{fig:X-ray}, the points on the surface of the NS are described by the
azimuthal angle $\phi_{\rm h}$ and polar angle $\theta_{\rm h}$. A photon
emitted at an angle $\theta_{\rm em}$ with respect to the surface normal escapes
to infinity at a different angle $\theta_{\rm h}$ due to the light bending 
effect in the curved spacetime. The relation between the two angles is given by
the ray tracing function~\citep{Pechenick:1983,Page:1994dz}
\begin{equation}
	\label{eqn:Theta_theta}
	\theta_{\rm h}(\theta_{\rm em})=\int_{0}^{R_{\mathrm{s}} / 2 R}  
	x\left[\left(1-\frac{R_{\mathrm{s}}}{R}\right)\left(\frac{R_{\mathrm{s}}}{2 R}\right)^{2}-(1-2 u) 
	u^{2} x^{2}\right]^{-1 / 2}\dd u\,,
\end{equation}
where $x\equiv \sin\theta_{\rm em}$, $R_{\rm s}\equiv 2GM/c^{2}$ is the
Schwarzschild radius.  In a flat spacetime, the visible condition is simply
$\cos\theta_{\rm h}>0$. The strong gravity of the NS allows the observer to see 
the region with negative $\cos\theta_{\rm h}$. The critical value of
$\cos\theta_{\rm h}$ that defines the dark side of star is determined by the
condition $\theta_{\rm em}=90^{\circ}$.

The differential spectral flux from the hotspot
is~\citep{Pechenick:1983,Beloborodov:2002mr}
\begin{align}
	\label{differential_flux}
	\dd F_{j}(E_{\infty}, \Phi)=&\left(1-\frac{r_{g}}{R}\right)^{1/2} I_{j}(\theta_{\rm em},E) \cos 
	\theta_{\rm em} \frac{\dd \cos \theta_{\rm em}}{\dd \cos \theta_{\rm h}} \frac{\dd S}{D^{2}}\nonumber\\
	=&\frac{R^{2}}{D^{2}} \left(1-\frac{r_{g}}{R}\right)^{1/2} I_{j}(\theta_{\rm em},E) \sin\theta_{\rm em} 
	\,\dd \sin \theta_{\rm em} \,\dd \phi_{\rm h}\nonumber\\
	=&\frac{R^{2}}{D^{2}} \left(1-\frac{r_{g}}{R}\right)^{1/2} I_{j}(\arcsin x,E) x \,\dd x \,\dd \phi_{\rm h}\,,
\end{align}  
where $\dd S=R^{2}\sin\theta_{\rm h}\dd \theta_{\rm h}\dd \phi_{\rm h}$ is the
visible surface element, $D$ is the distance between the NS and the observer,
$I_{j}\,(j=\rm X,O)$ are the specific intensities of the X and O mode photons at
the emission point, and $E_{\infty}=(1-R_{\rm s}/R)^{1/2}E$ is the observed
energy of the X-ray photons. The spectral flux then can be integrated
as~\citep{Page:1994dz}
\begin{equation}
	\label{eqn:integrated_flux}
	F_{j}(E_{\infty}, \Phi)=\frac{R^{2}}{D^{2}} \left(1-\frac{r_{g}}{R}\right)^{1/2}\int_{0}^{1}x I_{j}(\arcsin x,E)\dd x  
	\int_{0}^{2\pi}\dd \phi_{\rm h}\,.
\end{equation}
At a specific rotation phase $\Phi$, one first obtains the angle $\Theta$. Then
the ranges of $\theta_{\rm h}$ and $\phi_{\rm h}$ are determined by $\Theta$ and
the opening angle $\rho$. The dependence of $\theta_{\rm h}$ has been
transformed into that of $\theta_{\rm em}$ by the relation in
Eq.~(\ref{eqn:Theta_theta}). Finally, the observed spectral flux for a given
mode is given by Eq.~(\ref{eqn:integrated_flux}).  Note that the integration
domain is restricted to the hot region with the intensities being zero outside
the hot region. The observed spectral flux $F_{\rm I}$, $F_{\rm Q}$, and $F_{\rm
U}$ that are associated with the Stokes parameters $I$, $Q$, and $U$
are~\citep{vanAdelsberg:2006uu,vanAdelsberg:2009qj}
\begin{align}
	F_{\rm I}=&F_{\rm O}+F_{\rm X}\,,\\
	F_{\rm Q}=&F_{\rm I}\,\Pi_{\rm em}\cos 2\Psi(r_{\rm pl})\,,\\
	F_{\rm U}=& F_{\rm I}\,\Pi_{\rm em}\sin 2\Psi(r_{\rm pl})\,,
\end{align}
and the observed polarization fraction is 
\begin{equation}
	\Pi_{\rm L}= \frac{(F_{\rm Q}^{2}+F_{\rm U}^{2})^{1/2}}{F_{\rm I}}=\lvert \Pi_{\rm em}\rvert\,.
\end{equation}
The observed polarization fraction is equal to the intrinsic polarization
fraction, which arises from the assumption that the magnetic field is constant
across the hot region which is much smaller than the surface area of the star.

\begin{figure}
	\centering
	\includegraphics[width=8cm]{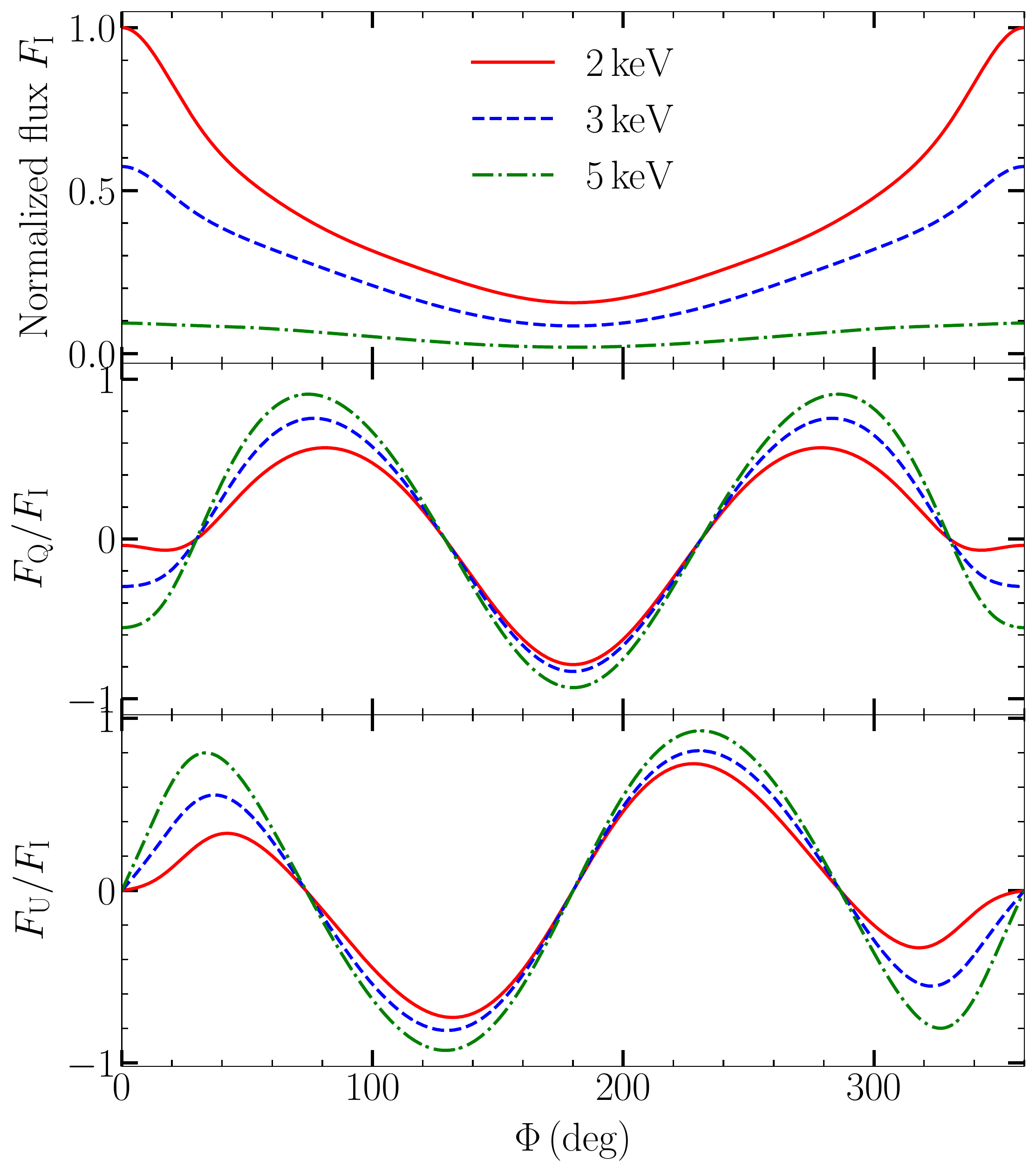}
	\caption{The phase evolutions of the spectral flux $F_{\rm I}$ (upper), the
	linear polarization $F_{\rm Q}/F_{\rm I}$ (middle), and the linear polarization
	$F_{\rm U}/F_{\rm I}$ (lower) for photon energies $E=2\,\rm keV, 3\,\rm
	keV$, and $5\,\rm keV$. The parameters of the model are the opening angle of
	the polar cap, $\rho=5^{\circ}$, the dipole magnetic field $B=10^{14}\,\rm
	G$, the effective temperature $T_{\rm eff}=5\times 10^{6}\,\rm K$, the
	inclination angle of the observer $\iota=45^{\circ}$, and the magnetic
	inclination angle $\alpha=65^{\circ}$.}
	\label{fig:stokes1}
\end{figure}

We give an example with a magnetic field $B=10^{14}\,\rm G$, an effective
temperature $T_{\rm eff}=5\times 10^{6}\,\rm K$, and a surface gravity $g=G
M/R^{2}\left(1-2 G M/R/c^{2}\right)^{-1 / 2}=2.4 \times 10^{14} \,\rm cm
\,s^{-2}$ in Fig.~\ref{fig:stokes1}.  The normalized $F_{\rm I}$, $F_{\rm
Q}/F_{\rm I}$, and $F_{\rm U}/F_{\rm I}$ for different photon energies are
shown. There are distinctive features that reflect the interplay between the NS
geometry, the strong magnetic field, and vacuum birefringence. Because the
magnetic field $B>B_{ l}\simeq 7\times 10^{13}\,\rm G$, the vacuum resonance
density lies between the decoupling densities of the X mode and O mode photons
($\rho_{\rm O}<\rho_{\rm V}<\rho_{\rm X}$), the linear polarization $F_{\rm
Q}/F_{\rm I}$ for different photon energies coincides in phase as the star
rotates~\citep{vanAdelsberg:2006uu}. 

The emergent radiation is dominated by the X mode except for $\theta_{\rm em}$
close to zero. The polarization degree $\Pi_{\rm L}$ is smaller when the rotation
phase $\Phi$ is close to 0 than for when it is close to $90^{\circ}$. This can be
understood by considering the variation in X and O mode opacities when varying the angle
between the photon propagation and magnetic field directions. In our chosen NS geometry, the emission angle $\theta_{\rm em}$ is closer to zero for a rotation
phase around $\sim 0^{\circ}$ compared to other phases. The difference between the X
and the O mode opacities becomes smaller. Thus, the polarization fraction is
smaller than at other phases.  In contrast, the emission angle $\theta_{\rm em}$ is
close to $\sim 45^{\circ}$ when the rotation phase is far away from $0^\circ$ or
$360^{\circ}$. At those angles, the difference between the X and O mode
opacities is maximal and the polarization fraction is larger.

\subsection{Modulations on the Stokes parameters of X-rays}

\begin{figure}
	\centering
	\includegraphics[width=8cm]{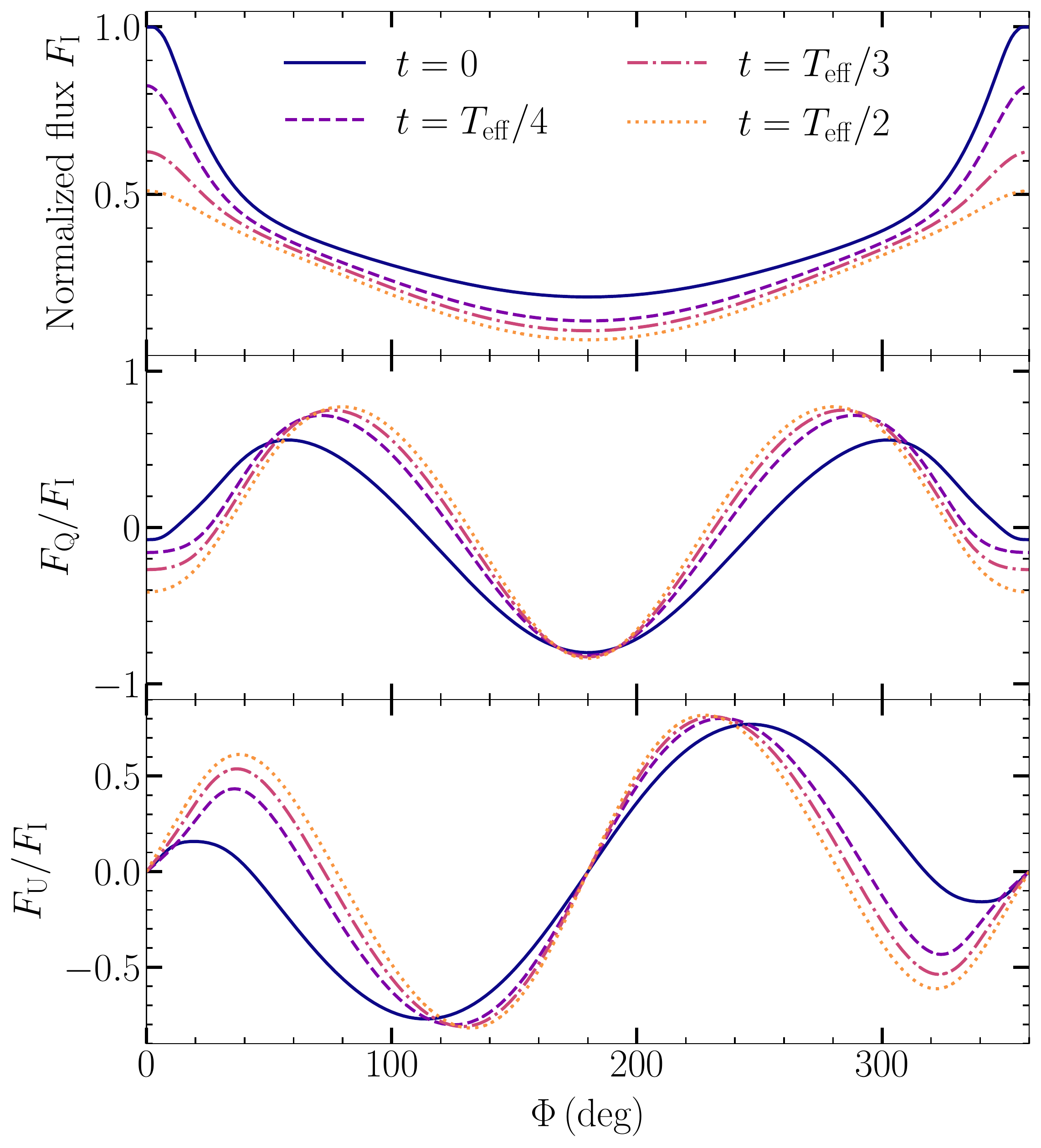}
	\caption{The phase-resolved spectral flux $F_{\rm I}$ (upper), $F_{\rm
	Q}/F_{\rm I}$ (middle), and $F_{\rm U}/F_{\rm I}$ (lower) of $E=3\,\rm keV$ 
	at different precession phases for a biaxial NS. The parameters are shown in 
	Case \uppercase\expandafter{\romannumeral1} of Table~{\ref{tab:third}}.}
	\label{fig:stokes2}
\end{figure}

\def\arraystretch{1.1}
\begin{table*}
	\caption{The intrinsic and effective parameters for modulated polarizations of X-rays and radio signals shown in Fig.~\ref{fig:stokes2}-\ref{fig:compare}.}
	\centering 
	\begin{tabular}{lcccccccccccccc}
		\toprule
		 Case&\multicolumn{8}{c}{Intrinsic parameters}&\multicolumn{6}{c}{Effective parameters} \\
		 \cmidrule(lr){2-9}\cmidrule(lr){10-15}
		 &
		$P_{0}\,(\rm s)$&
		$B\,(\rm G)$ & 
		$\epsilon$ & 
		$\delta$ & 
		$\theta_{0}\, (\degr)$ & 
		$\chi\, (\degr)$ & 
		$\eta \, (\degr) $&
		$\iota\,(\degr)$&
		$\epsilon_{\rm eff}$ &
		$\delta_{\rm eff}$ & 
		$\theta_{\rm eff, 0}\, (\degr)$ & 
		$\chi_{\rm eff}\, (\degr) $ & 
		$\eta_{\rm eff} \, (\degr) $ & 
		$T_{\rm eff}\,(\rm yr)$ \\ 
		\hline
		\uppercase\expandafter{\romannumeral1} & 5 & $ 10^{14}$ & $10^{-7}$ & 0 & 15&  65 & 0 & 45 & $1.01\times 10^{-7}$& 0.0124& 15.3&65.3 & 0 & 1.64\\
	
		\uppercase\expandafter{\romannumeral2} & 5 & $10^{14}$ & $10^{-7}$ & 1 & 15&  65 & 45 & 45 & $1.00\times 10^{-7}$& 0.993& 15.2 & 65.5 & 44.4 & 2.35 \\

		\uppercase\expandafter{\romannumeral3} & 5 & $ 10^{14}$ & $10^{-7}$ & 0 & 15&  45 & 0 & 45& $1.00\times 10^{-7}$& 0.00761& 15.4&45.4 & 0 & 1.65 \\
	
		\uppercase\expandafter{\romannumeral4} & 5 & $ 10^{14}$ & $10^{-7}$ & 1 & 15&  45 & 45 &  45 &$9.96\times 10^{-8}$& 1.01& 15.3 & 45.6 & 44.8 & 2.38 \\

		\uppercase\expandafter{\romannumeral5} & 5 & $5\times 10^{14}$ & $10^{-7}$ & 0 & 15&  45 & 0 &45 & $1.07\times 10^{-7}$& 0.261& 25.3&55.3 & 0 & 1.87 \\

		\uppercase\expandafter{\romannumeral6} & 5 & $5\times 10^{14}$ & $10^{-7}$ & 1 & 15&  45 & 45 &45 & $9.99\times 10^{-8}$& 1.15& 24.9&60.2 & 36.1 & 2.75 \\

		\uppercase\expandafter{\romannumeral7} & {5} & ${10^{14}}$ & ${10^{-5}}$ & {0} & {18}&  {40} & {0} &{45} & ${1.00\times 10^{-5}}$& ${6.20\times 10^{-5}}$& {18.0}&{40.0}&{0} & {0.0167}\\

		\uppercase\expandafter{\romannumeral8} &{5} & ${10^{14}}$ & ${10^{-5}}$ & {5}& {18}&  {40} & {0} &{45} & ${1.00\times 10^{-5}}$& ${5.00}$& {18.0}&{40.0}& {0} & {0.0490} \\
		
		\bottomrule
	\end{tabular}
	\label{tab:third}
\end{table*}


Since the phase evolution of the Stokes parameters are similar for different
energies in Fig.~\ref{fig:stokes1}, we fix $E=3\,\rm keV$ to study the
modulations due to the precession. In Fig.~\ref{fig:stokes2}, we show the phase
evolution of the normalized $F_{\rm I}$, $F_{\rm Q}/F_{\rm I}$, and $F_{\rm
U}/F_{\rm I}$ at different precession phases for a biaxial NS. Only half of the
precession period is shown because the precession is periodic. The modulations
for the triaxial case are similar. The precession mainly causes variations at
a rotation phase close to $0^{\circ}$. 

\begin{figure*}
	\centering
	\includegraphics[width=15cm]{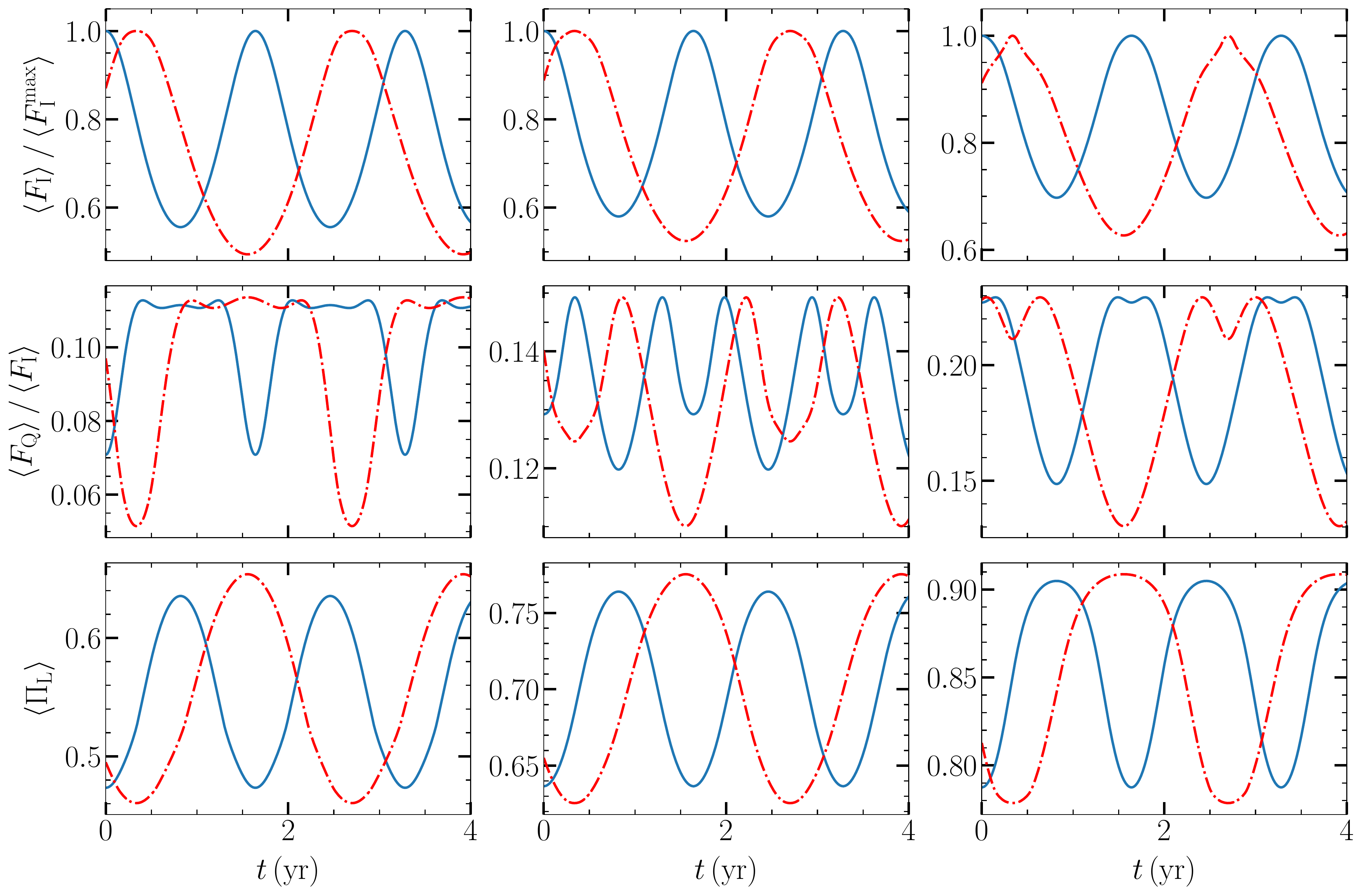}
	\caption{The time evolutions of the phase-averaged normalized spectral flux
	$\left\langle F_{\rm I}\right\rangle/\left\langle F_{\rm I}^{\rm
	max}\right\rangle$ (upper), the phase-averaged linear polarization
	$\left\langle F_{\rm Q} \right\rangle/\left\langle F_{\rm I}\right\rangle$
	(middle), and the phase-averaged polarization fraction $\left\langle
	\Pi_{\rm L}\right\rangle$ (lower) for photon energies $E=2\,\rm keV$ (left),
	$E=3\,\rm keV$ (middle) and $E=5\,\rm keV$ (right). The parameters for the biaxial and triaxial 
	cases are shown in Case \uppercase\expandafter{\romannumeral1} and Case 
	\uppercase\expandafter{\romannumeral2} of Table~{\ref{tab:third}} respectively.}
	\label{fig:stokes3}
\end{figure*}

The phase-resolved X-ray polarization is usually hard to get from observations.
Thus, we give the phase-averaged Stokes parameters $\left\langle F_{\rm
I}\right\rangle$, $\left\langle F_{\rm Q}\right\rangle/\left\langle F_{\rm
I}\right\rangle$ and polarization degree $\left\langle \Pi_{\rm L}\right\rangle$
in Fig.~\ref{fig:stokes3}. The phase averaged $F_{\rm U}$ is zero and it is
omitted in the figure.  Both biaxial and triaxial cases are shown and the
phase-averaged spectral flux $\left\langle F_{\rm I}\right\rangle$ is normalized
to the maximal value. 

The amplitude of $\left\langle F_{\rm I}\right\rangle/\left\langle F_{\rm
I}^{\rm max}\right\rangle$ can vary up to $\sim 40\%$ during the precession,
which is quite substantial. \citet{heyl2002hot} used free precession to explain
the flux variations from the magnetar 1E 161348$-$5055. They modeled the
hotspot emission in a similar way to our work. The large variations of the
phase-averaged flux are partially caused by the emission model. We assume that
the emission comes from a small hot region centered around the magnetic axis. If
the emission comes from different patches or even the whole stellar surface with
temperature profiles, the large modulations on the spectral flux might be
reduced.

The phase-averaged linear polarization $\left\langle F_{\rm Q}
\right\rangle/\left\langle F_{\rm I}\right\rangle$ and the phase-averaged 
polarization fraction $\left\langle \Pi_{\rm L}\right\rangle$ vary $\sim
10\%$--$20\%$ in our examples. Different from the spectral flux, the modulations
on the polarizations may not be reduced if the emission comes from different
patches of the stellar surface.  As we discussed before, when an X-ray photon
propagates in the magnetosphere, its polarization state evolves adiabatically
along the varying magnetic field up to the polarization limiting radius $r_{\rm
pl}$, which is far from the surface of the NS. Polarization states of photons 
from different patches of the star largely do not cancel. The magnetic field
direction at $r_{\rm pl}$ changes during the precession and the modulations
should always exist.

\subsection{Modulations on polarized radio emission}

\begin{figure}
	\centering
	\includegraphics[width=8cm]{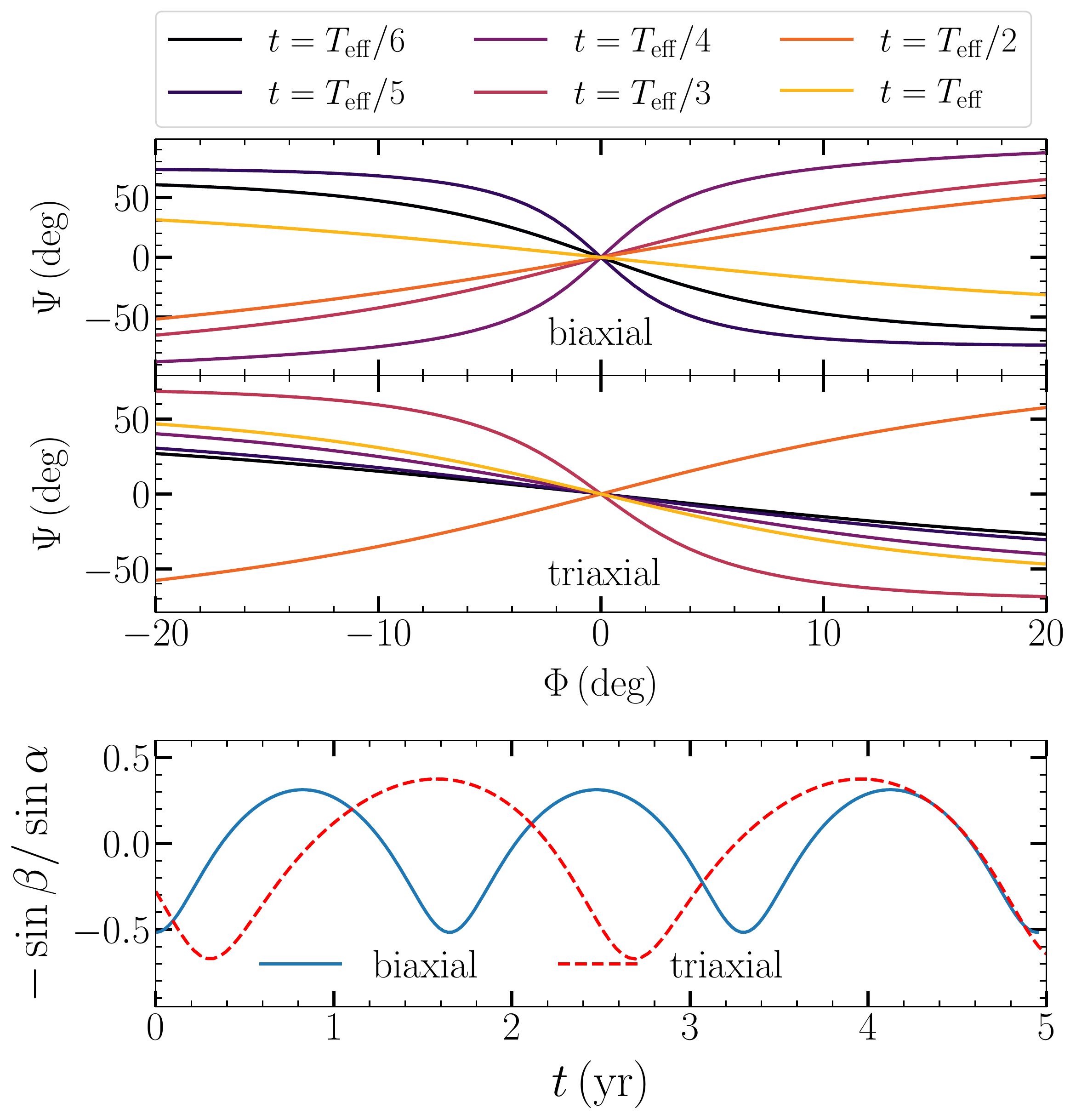}
	\caption{The upper panel shows the PA evolution at different precession
	phases for both biaxial and triaxial cases.  The lower panel shows the time
	evolution of the inverse of the steepest gradient, $-\sin\beta/\sin\alpha$.
	The parameters for the biaxial and triaxial 
	cases are shown in Case \uppercase\expandafter{\romannumeral3} and Case 
	\uppercase\expandafter{\romannumeral4} of Table~{\ref{tab:third}} respectively.}
	\label{fig:rvm1}
\end{figure}

\begin{figure}
	\centering
	\includegraphics[width=8cm]{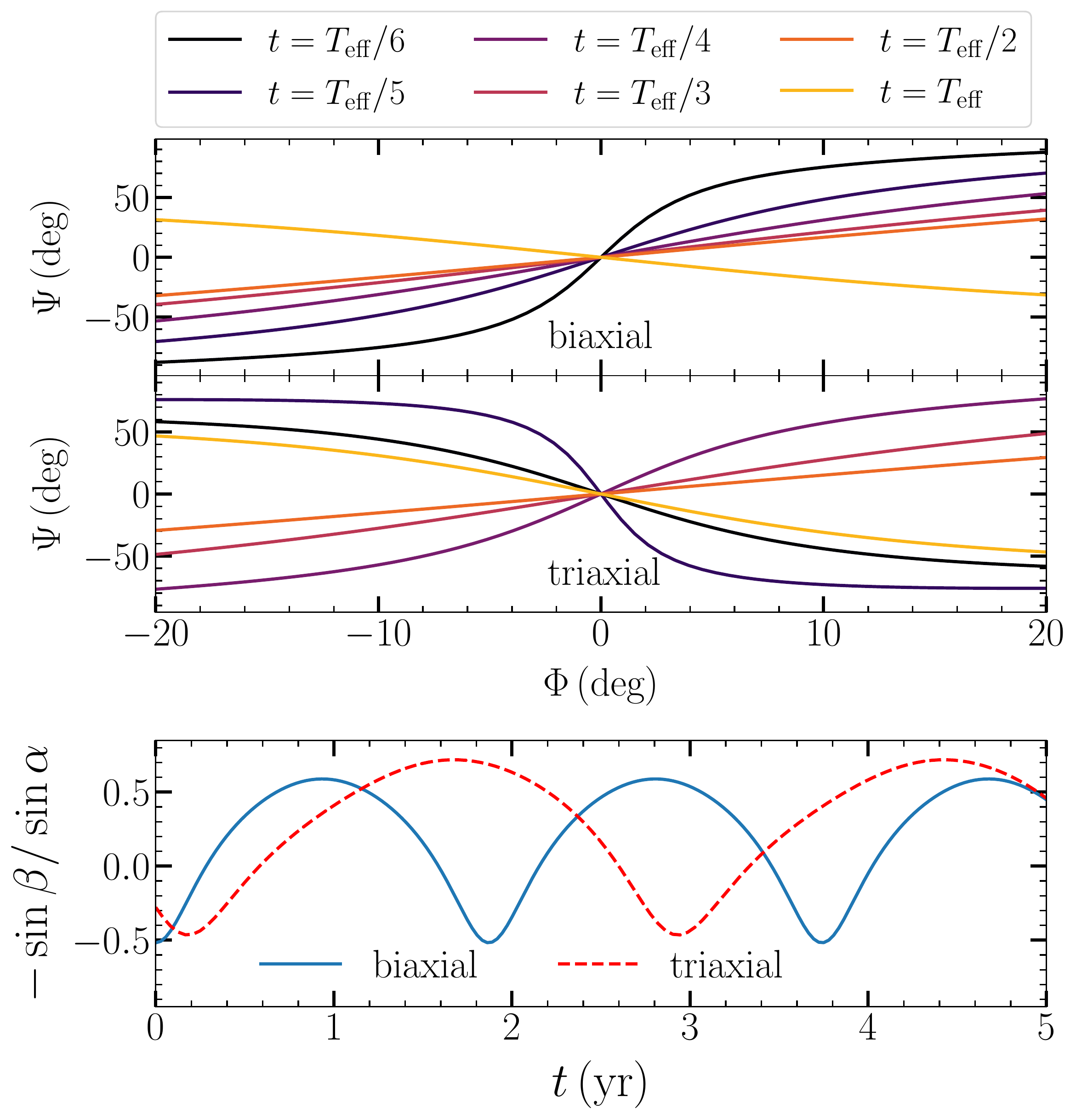}
	\caption{Same as Fig.~\ref{fig:rvm1} but with a larger magnetic field $B=5\times 10^{14}\,\rm
	G$. The parameters for the biaxial and triaxial 
	cases are shown in Case \uppercase\expandafter{\romannumeral5} and Case 
	\uppercase\expandafter{\romannumeral6} of Table~{\ref{tab:third}} respectively.}
	\label{fig:rvm2}
\end{figure}

\begin{figure}
	\centering
	\includegraphics[width=8cm]{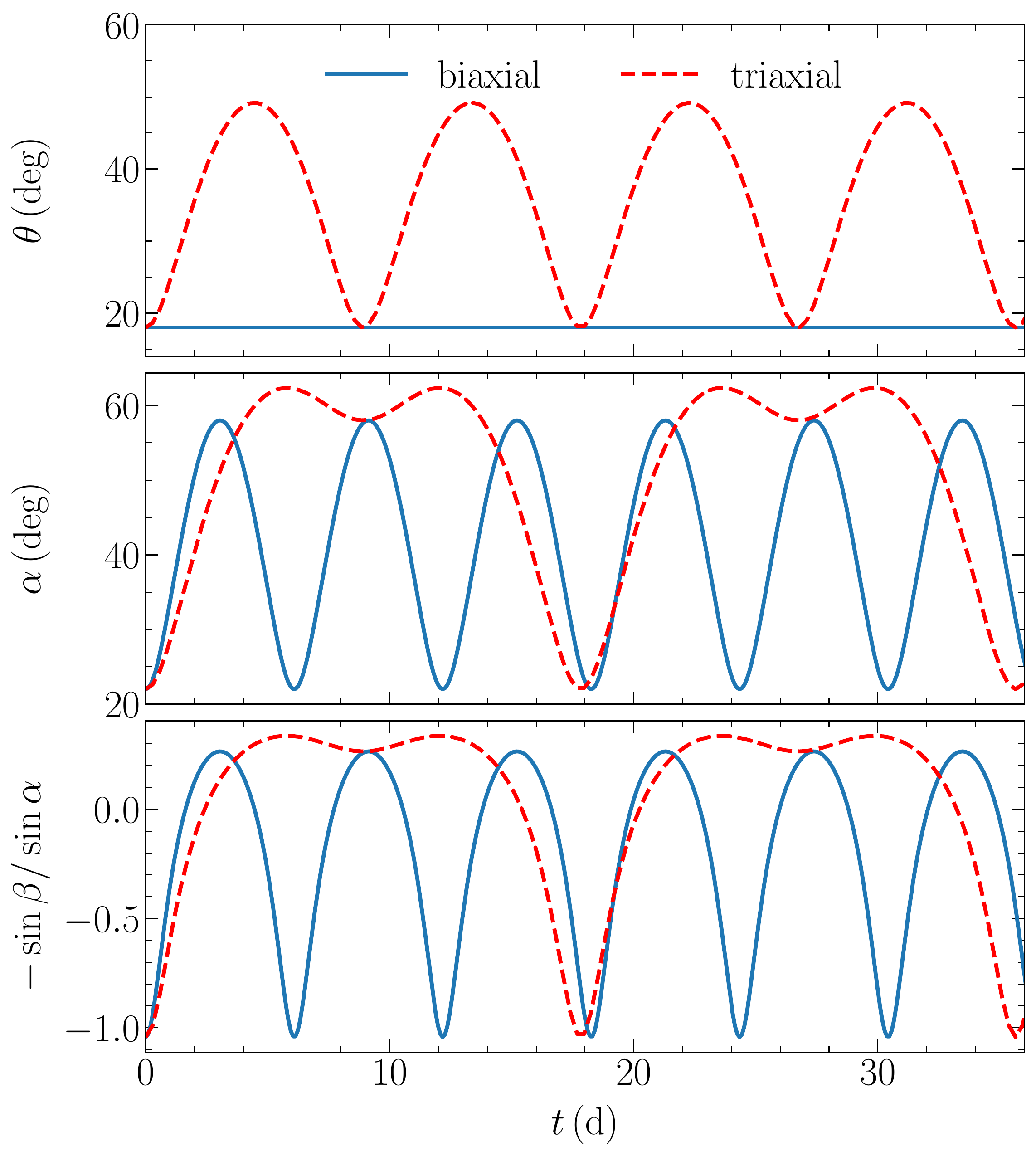}
	\caption{The time evolution of $\theta$, $\alpha$, and $-\sin\beta/\sin\alpha$. 
	The parameters for the biaxial and triaxial 
	cases are shown in Case \uppercase\expandafter{\romannumeral7} and Case 
	\uppercase\expandafter{\romannumeral8} of Table~{\ref{tab:third}} respectively.}
	\label{fig:compare}
\end{figure}

We use the RVM to study the modulations on the polarized radio emission.
It may be possible to observe the swing of the emission region and the
modulations on the PA due to the precession. 

We present the PA evolutions at different precession phases for a magnetar with
$B=10^{14}\,\rm G$ in the upper panel of Fig.~\ref{fig:rvm1}. Due to the
precession, the slope of the PA will change. The steepest gradient of the PA is
\begin{equation}
 \frac{\dd \Psi}{\dd \Phi} \bigg|_{\Phi=0}=-\frac{\sin\alpha}{\sin\beta}\,.
\end{equation}
Practically, the precession of magnetars may be observed from the variations of
the steepest gradient of the PA.  As we take the inclination angle of the
observer to be $\iota={\pi}/{4}$, the impact parameter $\beta=\iota-\alpha$
changes sign during precession. Thus the slope of the PA can potentially change sign. In
the lower panel of Fig.~\ref{fig:rvm1}, we show the inverse of the steepest
gradient, $-\sin\beta/\sin\alpha$ for a better
illustration. 

For comparison, we also give examples with $B=5\times 10^{14}\,\rm G$ in
Fig.~\ref{fig:rvm2}. The effects of the near-field torque cannot be neglected.
The wobble angle varies across a much larger range and the modulations of the
steepest gradient are distinct from that in Fig.~\ref{fig:rvm1}. Moreover, the 
differences between the biaxial and triaxial cases are more obvious than Fig.~\ref{fig:rvm1}.

{The parameter $m$ is on the order of $0.1$ for the examples shown in Table~\ref{tab:third}.
As shown in the lower panels of Fig.~\ref{fig:rvm1} and Fig.~\ref{fig:rvm2}, the modulations for 
the biaxial and triaxial cases are similar. \revise{In contrast}, the triaxiality can be 
observed directly from polarizations if $m$ is large enough. We show a triaxial case 
with $m=0.528$ in Fig.~\ref{fig:compare}. For the biaxial case, the wobble angle is constant, 
the variation of $\alpha$ and $-\sin\beta/\sin\alpha$ is harmonic and has only 
one peak. In fact, these features are true for any biaxial case according to Eq.~(\ref{eqn:biaxial_alpha}).
For the triaxial case, the modulations of $\alpha$ and $-\sin\beta/\sin\alpha$ are not 
harmonic. Due to the variation of the wobble angle, the time evolution of $\alpha$ and $-\sin\beta/\sin\alpha$
also shows a ``double-peak'' structure in our case.}

\section{Discussions}
\label{sec:discussion}

In our work, we gave the analytical solutions to the free precession of
triaxially-deformed NSs following~\citet{landau1960course},
\citet{Akgun:2005nd}, and \citet{Wasserman:2021dlh}. 
\revise{We assumed that the rotation is rigid and ignored superfluid pinning or
any internal dissipations.} The pinning of the superfluid can lead
to fast precession comparable to the spin frequency~\citep{Shaham1977,Sedrakian:1998vi}. 
It is also possible that the precession is
damped quickly by the coupling of normal fluids to the superfluid core if the
strong internal magnetic field of magnetars unpairs the proton superconductor in
the stellar core \citep{Sedrakian:2016vhm}. \revise{In this regard,} our studies can serve as a starting 
point to further investigate the internal couplings and dissipative mechanisms.

The strong magnetic fields of magnetars also induce large electromagnetic
torques, which are important to determine the spin and geometry evolutions of
precessing magnetars. Assuming a dipole field, we considered both the near-field
and far-field electromagnetic torques in the forced precession problem. 

For magnetars with large external magnetic fields, the near-field torque couples
to the precession solution and affects the motions substantially. This is the
central idea in the so called radiative precession advocated
by~\citet{Melatos1997,Melatos:2000qt}. The near-field torque can be effectively
absorbed into the moment of inertia tensor of the
star~\citep{Melatos:2000qt,Glampedakis:2010qm,Zanazzi:2015ida}. We solved the
forced precession problem analytically by transforming the near-field torque
into an effective deformation along the magnetic axis. We found that the effects
of the near-field torque cannot be ignored in the dynamical evolution if
$\epsilon_{\rm m}\gtrsim 0.1\epsilon$, where $\epsilon_{\rm m}$ is the effective 
ellipticity induced by the external magnetic field and $\epsilon$ is the ellipticity 
sourced from the magnetic and elastic deformations.

{The far-field torque leads to the spin down and secular change of the magnetic
inclination angle. Perturbation methods were used to study the forced precession under the far-field
torque. We obtained analytical solutions of the spin evolution for general
triaxial stars.} One part of the far-field torque comes from the direct
emission of electromagnetic waves due to the time-varying magnetic multipolar
moments of the star. Another part is caused by electromagnetic emission from
charged particles being accelerated in the magnetosphere. Therefore, we not only
used the simple vacuum torque, but also applied a parametrized plasma-filled
torque proposed by~\citet{Li:2011zh} and \citet{Philippov:2013aha} according to
MHD simulations. The form of the plasma-filled torque is equivalent to adding an
alignment component of the far-field torque compared to the vacuum case. 
Note that the near-field torque affects the spin-down rate indirectly
because it leads to the variations of the magnetic inclination angle for a
precessing magnetar. 

In our calculations, we assumed that the external magnetic field was dipolar. It
is commonly believed that higher multipoles should be considered
crucially for magnetars. \citet{Zanazzi:2015ida} investigated the near-field
torque contributed by the quadrupole field. The effective deformation is not
symmetric about a specific axis and can be classified into two independent 
components. The multipoles contributing to the near-field torque can also be
absorbed into the moment of inertia tensor of the star.  The solutions in our
work can still be applied but with different effective parameters. The direct
electromagnetic emission from the magnetic multipoles is probably dominated by
the magnetic dipole. On the perspective of observations, we may leave the
coefficients in the parametrized far-field torque in Eq.~(\ref{eqn:spindown}) as
free parameters to absorb the effects of higher magnetic multipoles, as well as
complex charges and currents in the magnetosphere.

During the precession, the torques are locked in phase with the precession,
which in turn modulates the spin-down rate.  We first studied the spin evolution
of triaxially-deformed magnetars and gave the analytical timing residuals, which
contained the geometric term resulting from the phase modulations of the
precession and the spin-down term arising from the varying far-field torque.

The polarization maps out the geometry of the emission region and can serve as a
useful probe to find the precession of magnetars. For the soft X-rays, we used
the model given in~\citet{Lai:2003nd} and \citet{vanAdelsberg:2006uu} to
calculate the observed Stokes parameters emitted from a hot region centered
around the magnetic dipole. The general relativistic effects and the vacuum
birefringence were considered. We investigated the modulations on the spectral
intensities, the linear polarization, and the polarization fraction for both
phase-resolved and phase-averaged scenarios. 
{For radio signals, we simply used the RVM to study the PA evolution
for different geometries during the precession.  It is possible to detect the
precession of transient magnetars through the variations of the steepest gradient
of the PA if large-amplitude precession is excited.} 

The polarization state of X-rays evolves adiabatically following the varying
magnetic field it experiences up to the polarization limiting radius $r_{\rm
pl}$. The polarization states of photons from different patches of the star tend
to align at $r_{\rm pl}$, and largely do not
cancel~\citep{Heyl:2003kt,Lai:2003nd}. Therefore, the polarizations can still be
modulated in the precession, even when the photons come from different patches (or even
the whole stellar surface) because the inclination of the magnetic field at
$r_{\rm pl}$ changes. In contrast, the modulations of the flux and the spectrum 
may be reduced or even eliminated if the emission comes from a large extended region. 

From the perspective of observations, IXPE has conducted the first
observation of the polarized X-ray emissions from the magnetar 
4U~0142+61~\citep{2022arXiv220508898T}. In near future, the eXTP mission will
give more accurate measurements of X-ray
polarizations~\citep{eXTP:2018anb, eXTP:2018kws} which will give us more
opportunities to find the precession of magnetars via polarizations.

\section{Conclusions}
\label{sec:conclusion}

{We gave a detailed model of precessing magnetars with triaxial deformation. 
The dynamical motion of the precession both in free and forced conditions was studied analytically.
For magnetars with $B\sim 5\times 10^{14}-10^{15}\,\rm G$, 
the effects of the electromagnetic torques must be considered crucially if the 
ellipticity $\epsilon \lesssim 10^{-7}$.} 

{Precession can produce observational consequences in timing and polarization. 
We gave the timing residuals from both the geometric term arising from the precession 
and the spin-down term arising from the variations of the far-field torque. 
The relative strength of the two terms is determined by the
relative strength between the rotation period $P$, the precession timescale
$\tau_{\rm p}$, and the spin-down timescale $\tau_{\rm rad}$. If $\tau_{\rm
p}/\tau_{\rm rad}\gg P/\tau_{\rm p}$, the spin-down term dominates. Otherwise,
the geometric term dominates over the spin-down term. }

{We also modeled the modulations on polarized X-ray and radio signals in different NS geometries.
Assuming the emission is centered around one of the magnetic pole, we showed that the 
prospects of detecting precession with polarization are promising
if large wobble angle is excited. Thanks to the QED effects in strongly magnetized magnetosphere, the 
modulations on the polarization of X-rays may always exist even if the emission comes from a much extended 
region or the whole star. Advanced detectors, such as IXPE and eXTP, will give us more
opportunities to find the precession of magnetars via polarizations.}

{A firm detection of magnetar precession will answer many questions about the strong internal 
magnetic field, the emission geometry, and the equation of state of NSs. 
Our timing and polarization models can be used to search and interpret magnetar precession.}

{In this work, we assume that the emission comes from a small region centered around the magnetic pole and the 
magnetic field itself is dipolar.
\revise{However}, the emission may come from a much extended region and possibly is distorted by the scattering
processes~\citep{Caiazzo:2021ode} for magnetars. The actual magnetic structures of magnetars 
are likely to have a twisted magnetic field configuration which contains
contributions from higher-order multipoles, affecting the polarization-state 
evolution for both radio signals from transient magnetars~\citep{Tong:2021anx}
and X-rays~\citep{Fernandez:2011aa,Taverna:2015vpa}.
We leave the modelling of the polarized X-rays and radio emission from 
precessing magnetars with more complex emission mechanisms and magnetic fields into future studies.}

\section*{Acknowledgements}

We thank Kuo Liu for carefully reading the manuscript, 
and Jingyuan Deng, Zexin Hu, and Rui Xu for discussions.  This work was
supported by the National SKA Program of China (2020SKA0120300), the National
Natural Science Foundation of China (11975027, 11991053, 11721303), the Max
Planck Partner Group Program funded by the Max Planck Society, and the
High-performance Computing Platform of Peking University.
DIJ acknowledges support from the STFC via grant number ST/R00045X/1.


\section*{Data Availability}
 
The data underlying this paper will be shared on reasonable request to 
the corresponding authors.



\bibliographystyle{mnras}
\bibliography{refs} 



\appendix

\section{Jacobi elliptic functions}
\label{append:A}

{The Jacobi elliptic functions are standard forms of elliptic functions. The three basic functions 
are denoted as $\operatorname{cn}(\tau,m)$, $\operatorname{sn}(\tau,m)$, and $\operatorname{dn}(\tau,m)$, 
where $0 \leq m \leq 1$. They naturally arise from the following integral
\begin{equation}
	\tau=\int_0^s\frac{\dd t}{\sqrt{1-m \sin ^2 t}}\,,
\end{equation}
where $s={\rm am}(\tau,m)$ is called the Jacobi amplitude. Then, it follows 
\begin{align}
	\cos s =& \operatorname{cn}(\tau,m)\,,\nonumber\\
	\sin s =& \operatorname{sn}(\tau,m)\,,\nonumber\\
	\sqrt{1-m \sin^{2}s}=& \operatorname{dn}(\tau,m)\,.
\end{align}
The Jacobi elliptic functions are periodic with period $T=4K(m)$, where $K(m)$ 
is the complete elliptic integral of the first kind.}

\begin{figure}
	\centering
	\includegraphics[width=8cm]{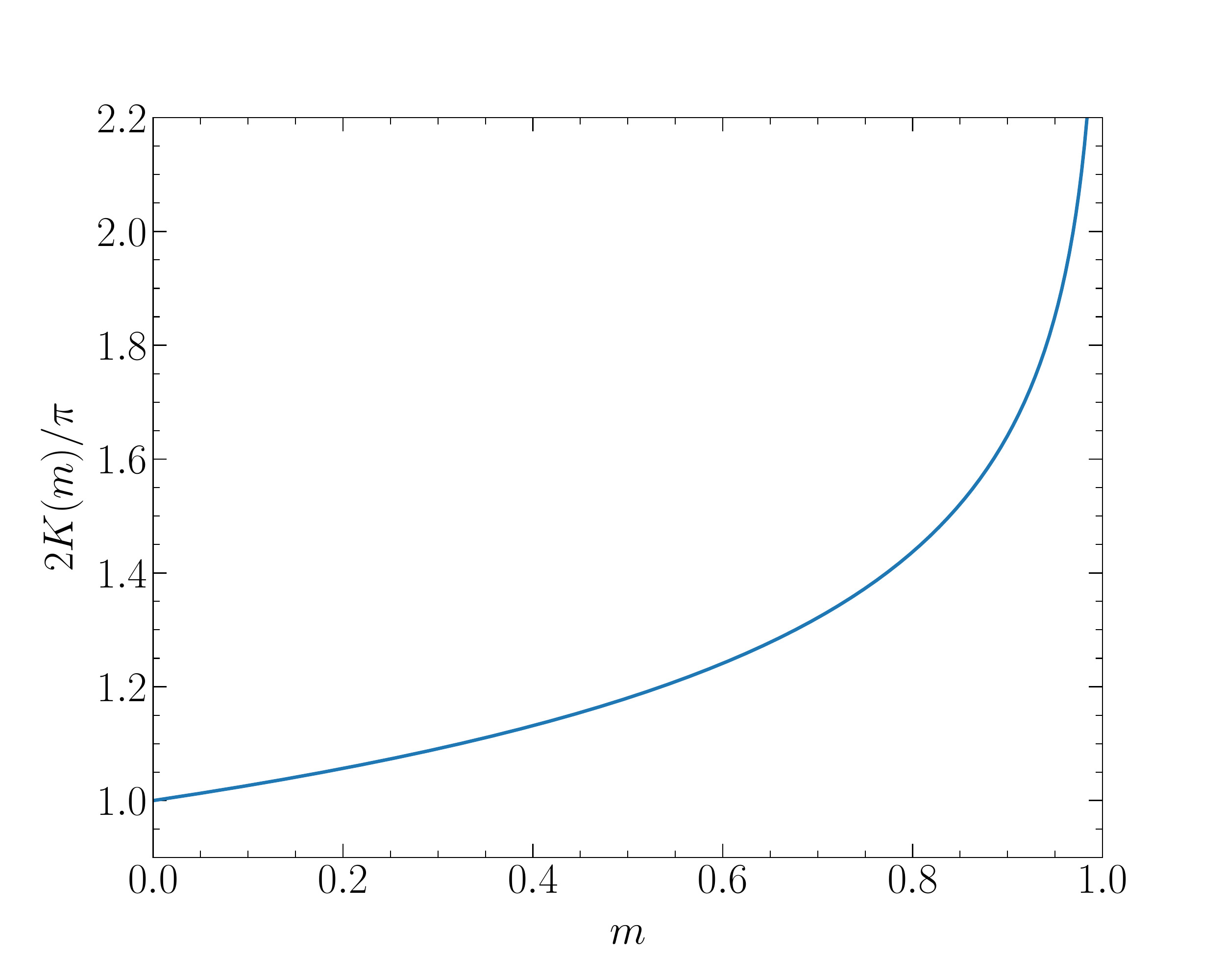}
	\caption{The relation between $m$ and $2K(m)/\pi$.}
	\label{fig:ellipk}
\end{figure}

{The expansions of the Jacobi elliptic function in series of $m$ are 
\begin{align}
	\operatorname{cn}(\tau,m)= \cos \tau -\frac{1}{8}m \sin \tau (-2\tau+\sin 2\tau)+\mathcal{O}(m^{2})\,,\nonumber\\ 
	\operatorname{sn}(\tau,m)= \sin \tau +\frac{1}{8}m \cos \tau (-2\tau+\sin 2\tau)+\mathcal{O}(m^{2})\,,\nonumber\\
	\operatorname{dn}(\tau,m)=1-\frac{1}{2}m\sin^{2}\tau +\mathcal{O}(m^{2})\,.
\end{align}
When $m=0$, $K(0)=\pi/2$, the variable $\tau$ equals to the Jacobi amplitude $s$, and the three elliptic functions turn
into 
\begin{align}
	\operatorname{cn}(\tau,0)&= \cos \tau\,,\nonumber\\
	\operatorname{sn}(\tau,0)&= \sin \tau\,,\nonumber\\
	\operatorname{dn}(\tau,0)&= 1\,.
\end{align}
The trigonometric functions are $2\pi$ periodic. 
In Fig.~\ref{fig:ellipk}, we show the relation between the parameter $m$ and the 
period of the Jacobi elliptic functions over the period of the trigonometric functions.}

{The quantity $\cos \alpha$ shown in Eq.~(\ref{eqn:cosalpha}) is a function of the Jacobi 
elliptic functions. The integration of $\cos^{2}\alpha$ is
\begin{align}
	\label{eqn:int_alpha}
	\int_{0}^{t}\cos^{2}\alpha \dd t=&\frac{\cos ^{2} \theta_{0}}{\omega_{\rm p}\delta}\left[\hat{\mu}_{1}^{2}-(1+\delta) 
	\hat{u}_{2}^{2}+ \hat{\mu}_{3}^{2}\delta\right] E\left({\rm a m} \, \tau\right)\nonumber \\
	&  + \frac{1}{\omega_{\rm p}}\left[-\sin 2 \theta_{0}(1+\delta)^{\frac{1}{2}} \hat{\mu}_{2} 
	\hat{\mu}_{3}\operatorname{cn}\tau\right]\nonumber\\
	& +  \frac{1}{\omega_{\rm p}}\sin 2\theta_{0} \hat{\mu}_{1} \hat{\mu}_{3} \operatorname{sn}\nonumber\\
	& + \frac{\cos ^{2} \theta_{0}}{\delta\omega_{\rm p}}\left[(-1+\delta \tan ^{2} \theta_{0}) \hat{\mu}_{1}^{2}+(1+\delta) 
	\hat{\mu}_{2}^{2}\right] \tau\nonumber\\
	& -\frac{2 \cos ^{2} \theta_{0} \hat{\mu}_{1} \hat{\mu}_{2}(1+\delta)^{\frac{1}{2}}}{\omega_{\rm p}\delta}\operatorname{dn}\tau 
	+A_{\rm c}\,,
\end{align}
where $E({\rm a m}\, \tau)$ is the Jacobi elliptic integral of the second kind,
and ${\rm am}\, \tau=\arcsin(\operatorname{sn}\tau)$ is the Jacobi amplitude.
The term $A_{\rm c}$ is an integration constant, which can be obtained directly by
setting the integral to be zero at the initial time $t=0$.}

\bsp	
\label{lastpage}
\end{document}